\documentclass[sn-mathphys-ay]{sn-jnl}

\usepackage{geometry}
\usepackage{graphicx}%
\usepackage{multirow}%
\usepackage{amsmath,amssymb,amsfonts}%
\usepackage{amsthm}%
\usepackage{mathrsfs}%
\usepackage[title]{appendix}%
\usepackage{xcolor}%
\usepackage{textcomp}%
\usepackage{manyfoot}%
\usepackage{booktabs}%
\usepackage{algorithm}%
\usepackage{algorithmicx}%
\usepackage{algpseudocode}%
\usepackage{listings}%
\graphicspath{{figures/}}
\usepackage{xurl}
\usepackage[size=footnotesize]{subcaption}
\usepackage{siunitx}
\usepackage{cleveref}
\usepackage{tikz}
\usetikzlibrary{shapes.geometric, arrows, positioning, fit}
\usepackage{blindtext}
\usepackage{verbatim}
\usepackage[colorinlistoftodos,prependcaption,textsize=tiny]{todonotes}

\newcommand\itblue[1]{\textcolor{blue}{\textit{#1}}}

\colorlet{Reviewer1}{orange}
\colorlet{Reviewer2}{cyan}

\theoremstyle{thmstyleone}%
\theoremstyle{thmstyletwo}%

\theoremstyle{thmstylethree}%

\begin{document}

\title[Article Title]{Experimental study of dynamic wetting behavior through curved microchannels with automated image analysis}


\author[1,3]{\fnm{Huijie} \sur{Zhang}}\email{Huijie.Zhang@bosch.com}

\author*[2]{\fnm{Anja} \sur{Lippert}}\email{Anja.Lippert@bosch.com}

\author[2]{\fnm{Ronny} \sur{Leonhardt}}\email{Ronny.Leonhardt@bosch.com}

\author[2]{\fnm{Tobias} \sur{Tolle}}\email{Tobias.Tolle@bosch.com}

\author[2,3]{\fnm{Luise} \sur{Nagel}}\email{Luise.Nagel@bosch.com}

\author[3]{\fnm{Mathis} \sur{Fricke}}\email{fricke@mma.tu-darmstadt.de}

\author[3]{\fnm{Tomislav} \sur{ Mari\'c}}\email{maric@mma.tu-darmstadt.de}

\affil*[1]{\orgdiv{Mobility Electronics}, \orgname{Robert Bosch GmbH}, \orgaddress{\street{Markwiesenstrasse 46}, \city{Reutlingen}, \postcode{72770}, \country{Germany}}}

\affil[2]{\orgdiv{Corporate Research}, \orgname{Robert Bosch GmbH}, \orgaddress{\street{Robert-Bosch-Campus 1}, \city{Renningen}, \postcode{71272}, \country{Germany}}}

\affil[3]{\orgdiv{Mathematical Modeling and Analysis}, \orgname{Technical University of Darmstadt}, \orgaddress{\street{Peter-Grünberg-Straße 10}, \city{Darmstadt}, \postcode{64287}, \country{Germany}}}


\abstract{\itblue{This is the author manuscript of an article \cite{zhang_published_2024} published in Experiments in Fluids, available at \url{https://rdcu.be/dJ84G}. Please cite the published journal article \cite{zhang_published_2024} when refering to this manuscript.} Preventing fluid penetration poses a challenging reliability concern in the context of power electronics, which is usually caused by unforeseen microfractures along the sealing joints. A better and more reliable product design heavily depends on the understanding of the dynamic wetting processes happening inside these complex microfractures, i.e. microchannels. A novel automated image processing procedure is proposed in this work for analyzing the moving interface and the dynamic contact angle in microchannels. In particular, the developed method is advantageous for experiments involving non-transparent samples, where extracting the fluid interface geometry poses a significant challenge. The developed method is validated with theoretical values and manual measurements and exhibits high accuracy. The implementation is made publicly available. The developed method is validated and applied to experimental investigations of forced wetting with two working fluids (water and 50 wt\% Glycerin/water mixture) in four distinct microchannels characterized by different dimensions and curvature. The comparison between the experimental results and molecular kinetic theory (MKT) reveals that the dynamic wetting behavior can be described well by MKT, even in highly curved microchannels. The dynamic wetting behavior shows a strong dependency on the channel geometry and curvature.}

\keywords{dynamic contact angle, molecular kinetic theory (MKT), forced wetting, curved microchannel, automated image analysis}



\maketitle







\section{Introduction}
\label{sec:01_introduction}

In recent years, with the automotive industry's strategy moving towards vehicle electrification, power electronics with higher energy density and efficiency have received considerable attention along with stringent reliability constrains. Reliability of power electronics involves multiple aspects, including mechanical strength, corrosion resistance and media tightness etc. (\cite{WANG6532474}; \cite{c1c7d6568ddf4fa38dd15078b692e6a9}). A highly challenging task is to prevent the fluid penetration caused by the unforeseen microfracture along sealings and faulty joints as reported by \cite{MALKIEL2023112159}. These microfractures exhibit complex characteristics which hinder an instinctive prediction of possible penetration rates, hence, making a deeper understanding of the dynamic wetting behavior, characterized by the dynamic contact angle, is necessary for reliable product design.

Accurate, robust and automatic measurement of dynamic contact angles is crucial for experimental and numerical investigations of wetting processes. In this work, a novel automated image analysis procedure is proposed for the interface detection and dynamic contact angle measurement of capillary flows. The automated image analysis is first validated in \ref{subsec:5.1_validation}, by comparing automatic measurements with theoretical values and manual measurements. The proposed automated procedure considerably speeds up experimental data analysis. An implementation of our procedure used in this manuscript is made publicly available at the TUDatalib data repository at TU Darmstadt (\cite{TUDatalib}), and at Bosch Research GitHub (\cite{automatedImageAnalysis}).

A robust, accurate, and automatic dynamic contact angle measurement provides a data for understanding dynamic wetting processes in capillaries. Dynamic wetting is very actively researched topic in interfacial science, combining analytical, experimental and numerical investigations, and generating a broad selection of models. Over the past decades, several theoretical approaches have been proposed to describe the dynamic wetting process and recent reviews where given by \cite{10.1063/5.0102028} and \cite{ZHANG2023102861}. The hydrodynamic theory (HDT) describes the contact line dynamics by focusing on the bulk motion of the liquid on the solid surface, which considers the viscous bending  of the liquid-free surface geometry in the mesoscopic region (\cite{Voinov1976HydrodynamicsOW}; \cite{Cox1986169}) as depicted in \cref{fig:2.2.1_HDT}. The viscous dissipation in the region close to the contact line is considered to be the dominant channel of dissipation in the HDT. As a quantitative approach, HDT is mostly used to interpret the wetting kinetics with small capillary number $Ca$ and small contact angles. The molecular-kinetic theory (MKT) (\cite{BLAKE1969421}; \cite{Blake1993}; \cite{BLAKE20061}) concentrates on the contact line friction dissipation in the vicinity of the three-phase contact zone. In essence, the dynamic wetting process is modeled as molecular adsorption and desorption process. The key parameters for the model are the equilibrium frequency of molecular displacements $\kappa^0$ and the average distance $\lambda$ of each displacement. Together with the temperature, these parameters determine the so-called contact line friction $\zeta$ (see details in \cref{subsec:02_MKT}). It is reported by \cite{doi:10.1021/acs.langmuir.6b00747} and \cite{MOHAMMADKARIM2018658} that MKT is appropriate to outline the forced movement of dynamic contact line on a solid surface with heterogeneity, e.g. chemical contamination or roughness. Considering the limitations and advantages of both models, a combined molecular-hydrodynamic model is developed by \cite{doi:10.1021/la00043a013} and \cite{BROCHARDWYART19921}, also \cite{YANG202021}. However, the existence of the free model parameters still requires reproducible and reliable experiments to serve as a robust basis for the curve-fitting. The Interface Formation Model (IFM) proposed by \cite{SHIKHMURZAEV1993589} introduces surface mass densities into the continuum mechanical model and describes the
dynamic wetting process through a mass transfer between bulk and surface layers. The dynamic contact angle is determined by a dynamic version of Young's equation where the surface tension depends on the local thermodynamic state of the interface. The model was successfully applied to various dynamic wetting phenomena (\cite{Shikhmurzaev2007CapillaryFW}). However, it is difficult to apply in practice because it contains a large number of unknown parameters. Another class of models in the framework of phase-field theories describes the interface as well as the contact line as a diffuse layer with a characteristic thickness (see e.g., \cite{van_de_waals}; \cite{Cahn}; \cite{Cahn_Hilliard}; \cite{YUE_ZHOU_FENG_2010}). Notably, in this modeling framework, the contact line is able to move by diffusion of the order parameter without the need for hydrodynamic slip at the solid boundary (\cite{JACQMIN_2000}). Moreover, the precursor-film model proposed by \cite{PhysRevLett.104.246101}, assuming the solid-liquid interface to contain a precursor-film (\cite{Hardy1919}), has the advantage of being persistent with HDT and suitable for contact line physics on heterogeneous surfaces.

For investigating the two mechanisms for contact line dynamics, spontaneous and forced wetting, extensive studies exist. Experimental setups for spontaneous wetting include droplet spreading (\cite{Eddi2013}; \cite{PhysRevFluids.2.043602}; \cite{doi:10.1021/acs.langmuir.7b01223}; \cite{doi:10.1021/acs.langmuir.6b01357}) and capillary rise (\cite{Qur1997InertialC}; \cite{SAUER199828}; \cite{doi:10.1021/la051341z}; \cite{clanet_quere_2002}), while the Wilhelmy plate (\cite{RAME1997245}; \cite{LU201643}; \cite{10.1063/5.0102028}) is a classical example for forced wetting. The forced wetting through microchannels, which holds considerable more relevance in the context of industrial applications, shall be the focus of this paper's investigation.

This work studies experimentally the dynamic wetting with small capillary number $Ca$ $(10^{-6} - 10^{-3})$ in  microchannels with different geometries  (\cref{sec:03_experiment}), with the intention of covering common industrial circumstances in the context of media tightness and leakage. Two working fluids, water and 50 wt\% Glycerin/water mixture are employed. The experimental results are compared with five different dynamic contact angle models (\cref{tab:2.1_dynCA_models}).

\section{Dynamic contact angle models}
\label{sec:02_dynamicContactAngle}

Contact line (CL) dynamics can be categorized into two types: spontaneous and forced. The movement of CL on a solid substrate (\cref{fig:2.2.1_HDT}) towards the equilibrium state is defined as the spontaneous motion, where the CL motion is dominated by the capillary force. As the CL moves, the dynamic contact angle (CA) $\theta_d$ and the CL velocity $U$ relax to reach the equilibrium state, with the aim of reducing $U$ to zero and reaching the equilibrium CA $\theta_e$. In the spontaneous category, the CL velocity $U$ is usually not constant.

The movement of the CL can be controlled by applying an external driving force as forced wetting/de-wetting, which could allow a constant CL velocity $U$. Different from the spontaneous mechanism, this forced motion is regarded as the force balance between the CL motion of achieving the equilibrium and the external force of pushing the CL to deviate from the equilibrium state. 

In addition to the CL velocity and static CA dependence of the dynamic CA, the CL motion is in strong relation to the surface roughness, chemical contamination, material pairing, temperature etc as reported by \cite{RevModPhys.57.827}. Numerous studies have been conducted to describe the CL dynamics theoretically, which essentially boil down to the two most influential models, namely hydrodynamic theory (HDT) and molecular kinetic theory (MKT). In this work, the experimental results discussed in \cref{sec:05_results} are compared with five dynamic CA models as listed in \cref{tab:2.1_dynCA_models}.

\subsection{Hydrodynamic Theory}
\label{subsec:01_HDT}

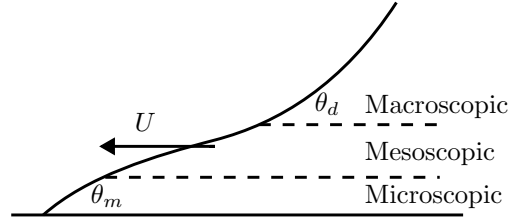
\begin{figure}[!htb]
    \centering
    \def\svgwidth{0.4\textwidth}
\begingroup%
  \makeatletter%
  \providecommand\color[2][]{%
    \errmessage{(Inkscape) Color is used for the text in Inkscape, but the package 'color.sty' is not loaded}%
    \renewcommand\color[2][]{}%
  }%
  \providecommand\transparent[1]{%
    \errmessage{(Inkscape) Transparency is used (non-zero) for the text in Inkscape, but the package 'transparent.sty' is not loaded}%
    \renewcommand\transparent[1]{}%
  }%
  \providecommand\rotatebox[2]{#2}%
  \newcommand*\fsize{\dimexpr\f@size pt\relax}%
  \newcommand*\lineheight[1]{\fontsize{\fsize}{#1\fsize}\selectfont}%
  \ifx\svgwidth\undefined%
    \setlength{\unitlength}{218.04901256bp}%
    \ifx\svgscale\undefined%
      \relax%
    \else%
      \setlength{\unitlength}{\unitlength * \real{\svgscale}}%
    \fi%
  \else%
    \setlength{\unitlength}{\svgwidth}%
  \fi%
  \global\let\svgwidth\undefined%
  \global\let\svgscale\undefined%
  \makeatother%
  \begin{picture}(1,0.47448123)%
    \lineheight{1}%
    \setlength\tabcolsep{0pt}%
    \put(0,0){\includegraphics[width=\unitlength,page=1]{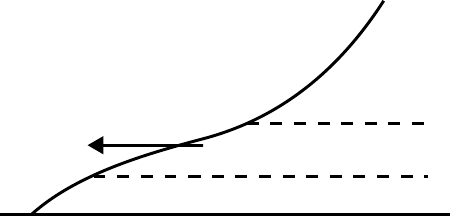}}%
    \put(0.17478676,0.03016971){\color[rgb]{0,0,0}\makebox(0,0)[lt]{\lineheight{1.25}\smash{\begin{tabular}[t]{l}$\theta_m$\end{tabular}}}}%
    \put(0.66557742,0.23299756){\color[rgb]{0,0,0}\makebox(0,0)[lt]{\lineheight{1.25}\smash{\begin{tabular}[t]{l}$\theta_d$\end{tabular}}}}%
    \put(0.27104064,0.18833938){\color[rgb]{0,0,0}\makebox(0,0)[lt]{\lineheight{1.25}\smash{\begin{tabular}[t]{l}$U$\end{tabular}}}}%
    \put(0.77572468,0.22818298){\color[rgb]{0,0,0}\makebox(0,0)[lt]{\lineheight{1.25}\smash{\begin{tabular}[t]{l}Macroscopic\end{tabular}}}}%
    \put(0.77572468,0.12363712){\color[rgb]{0,0,0}\makebox(0,0)[lt]{\lineheight{1.25}\smash{\begin{tabular}[t]{l}Mesoscopic\end{tabular}}}}%
    \put(0.77572468,0.02900216){\color[rgb]{0,0,0}\makebox(0,0)[lt]{\lineheight{1.25}\smash{\begin{tabular}[t]{l}Microscopic\end{tabular}}}}%
  \end{picture}%
\endgroup%

    \caption{Schematic illustration of the hydrodynamic theory.}
    \label{fig:2.2.1_HDT}
\end{figure}

The hydrodynamic theory (HDT) describes the physics of the fluid flow near the dynamic CL by dividing the moving CL region into three relevant length scales, namely microscopic, mesoscopic and macroscopic (\cite{Cox1986169}; \cite{Bonn2009739}; \cite{RAMIASA2014275}; \cite{LU201643}), as shown in \cref{fig:2.2.1_HDT}. The HDT considers the viscous dissipation due to the liquid flow close to the CL on the solid surface by applying the lubrication approximation on the Navier-Stokes equations. The hydrodynamic model derived by \cite{Voinov1976HydrodynamicsOW} and \cite{Cox1986169} is given as

\begin{equation}
    \theta_d^3 = \theta_0^3 + 9 \mathit{Ca} \ln \left(\frac{L}{L_s} \right).
    \label{eq:2.1_HDT}
\end{equation}

In \cref{eq:2.1_HDT}, $\theta_d$ represents the dynamic CA, which is the "apparent" CA measured at length scale $L$ and $\theta_m$ is the microscopic CA. It is assumed that $\theta_m=\theta_0$ with $\theta_0$ as the static CA (in practice, $\theta_0$ is usually equal to the equilibrium CA $\theta_e$). $\mathit{Ca}$ depicts the capillary number ($\mathit{Ca} = \mu U/ \sigma$), where $U$ is the CL velocity, $\mu$ is the dynamic viscosity of liquid and $\sigma$ is the surface tension force. $L$ is the characteristic length scale of the flow in the macroscopic region, while $L_s$ is the slip length. The slip length $L_s$ might be dependent on the surface roughness, shear rate, flow regime etc as reported by \cite{10.1063/1.3121305} and \cite{PhysRevE.81.011606}. Usually, the quantity $\ln \left(\frac{L}{L_s} \right)$ is treated as an adjustable parameter and can be fitted to experimental data.

\subsection{Molecular Kinetic Theory}
\label{subsec:02_MKT}

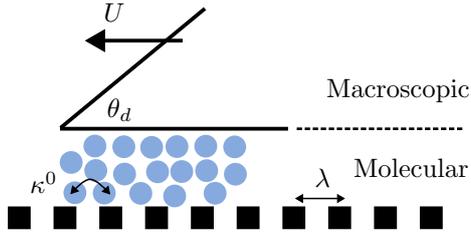
\begin{figure}[!htb]
    \centering
    \def\svgwidth{0.4\textwidth}
\begingroup%
  \makeatletter%
  \providecommand\color[2][]{%
    \errmessage{(Inkscape) Color is used for the text in Inkscape, but the package 'color.sty' is not loaded}%
    \renewcommand\color[2][]{}%
  }%
  \providecommand\transparent[1]{%
    \errmessage{(Inkscape) Transparency is used (non-zero) for the text in Inkscape, but the package 'transparent.sty' is not loaded}%
    \renewcommand\transparent[1]{}%
  }%
  \providecommand\rotatebox[2]{#2}%
  \newcommand*\fsize{\dimexpr\f@size pt\relax}%
  \newcommand*\lineheight[1]{\fontsize{\fsize}{#1\fsize}\selectfont}%
  \ifx\svgwidth\undefined%
    \setlength{\unitlength}{169.39910078bp}%
    \ifx\svgscale\undefined%
      \relax%
    \else%
      \setlength{\unitlength}{\unitlength * \real{\svgscale}}%
    \fi%
  \else%
    \setlength{\unitlength}{\svgwidth}%
  \fi%
  \global\let\svgwidth\undefined%
  \global\let\svgscale\undefined%
  \makeatother%
  \begin{picture}(1,0.50178145)%
    \lineheight{1}%
    \setlength\tabcolsep{0pt}%
    \put(0,0){\includegraphics[width=\unitlength,page=1]{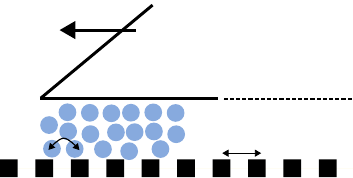}}%
    \put(0.69884777,0.29057401){\color[rgb]{0,0,0}\makebox(0,0)[lt]{\lineheight{1.25}\smash{\begin{tabular}[t]{l}Macroscopic\end{tabular}}}}%
    \put(0.75847175,0.11509146){\color[rgb]{0,0,0}\makebox(0,0)[lt]{\lineheight{1.25}\smash{\begin{tabular}[t]{l}Molecular\end{tabular}}}}%
    \put(0.21050356,0.45933317){\color[rgb]{0,0,0}\makebox(0,0)[lt]{\lineheight{1.25}\smash{\begin{tabular}[t]{l}$U$\end{tabular}}}}%
    \put(0.0486269,0.07938767){\color[rgb]{0,0,0}\makebox(0,0)[lt]{\lineheight{1.25}\smash{\begin{tabular}[t]{l}$\kappa^0$\end{tabular}}}}%
    \put(0.6736228,0.09315591){\color[rgb]{0,0,0}\makebox(0,0)[lt]{\lineheight{1.25}\smash{\begin{tabular}[t]{l}$\lambda$\end{tabular}}}}%
    \put(0.21811882,0.24851335){\color[rgb]{0,0,0}\makebox(0,0)[lt]{\lineheight{1.25}\smash{\begin{tabular}[t]{l}$\theta_d$\end{tabular}}}}%
  \end{picture}%
\endgroup%

    \caption{Schematic illustration of the molecular kinetic theory.}
    \label{fig:2.2.2_MKT}
\end{figure}

The MKT, proposed by \cite{BLAKE1969421}, concentrates on the molecular motion in the vicinity of the dynamic contact ``line" due to the friction dissipation, which can be balanced by the out-of-balance Young stress $F_f = \sigma(\cos{\theta_0}-\cos{\theta_d})$. Note here that on a molecular level, a ``line" defines a transition region (\cref{fig:2.2.2_MKT}). The solid surface is considered to contain many active absorption and desorption sites, where the molecules can attach and detach with the frequency $\kappa^0$ at equilibrium. The average distance between the active sites is denoted as $\lambda$ as displayed in \cref{fig:2.2.2_MKT}. Since it is assumed that the active sites are distributed uniformly, the number of the active sited per unit area on the solid surface $n$ can be determined as $n=\frac{1}{\lambda^2}$. The dynamic macroscopic CA described by MKT is given as

\begin{equation}
    \theta_m = \cos^{-1} \left( \cos{\theta_0} - \frac{2 k_B T}{\sigma \lambda^2} \sinh^{-1} \left( {\frac{U}{2 \kappa^0 \lambda}} \right) \right),
    \label{eq:2.2_MKT}
\end{equation}
where $k_B$ is the Boltzmann constant and $T$ stands for the temperature. It should be noted that the molecular motion frequency $\kappa^0$ and the distance $\lambda$ are treated as adjustable parameters and can be estimated via curve-fitting to the experimental data. Nevertheless, it is generally found that $\lambda$ is in the order of molecular dimensions for small molecules from  \AA \ to \SI{}{nm} and $\kappa^0$ varies within several orders of magnitude from \SI{}{kHz} to \SI{}{GHz} (\cite{doi:10.1021/la4017917}; \cite{SEDEV2015661}).

In the works of \cite{BlakeDiss} and \cite{BLAKE1969421}, introducing the coefficient of the CL friction as $\zeta$~($\SI{}{Pa \cdot s}$), \cref{eq:2.2_MKT} can be written as

\begin{equation}
    \theta_m = \cos^{-1} \left( \cos{\theta_0} - \frac{U \zeta}{\sigma} \right) ,
    \label{eq:2.3_MKT_linear}
\end{equation}
with $\zeta = k_B T/ \kappa^0 \lambda^3$ for small $\mathit{Ca}$.

\subsection{Combined Model}
\label{subsec:03_combinedModel}
Based on experiments, it is often observed that neither HDT nor MKT can interpret the observed behavior in a satisfactory way. \cite{doi:10.1021/la00043a013} proposed a combined molecular-hydrodynamic approach to describe the wetting kinetics by combining \cref{eq:2.1_HDT} and \cref{eq:2.2_MKT}

\begin{equation}
    \theta_d^3 = \theta_m^3 + 9 Ca \beta
    \label{eq:2.4_VoinovCox}
\end{equation}
with $\beta = \ln \left(\frac{L}{L_s} \right)$ and $\theta_m$ being calculated from \cref{eq:2.3_MKT_linear} for small $\mathit{Ca}$. As two adjustable parameters, $\beta$ and $\zeta$ can be obtained again via curve-fitting algorithms.

\subsection{Empirical Model}
\label{subsec:04_empiricalModel}
Apart from the above mentioned approaches, two more empirical models \cite{JIANG197974} and \cite{10.1007/BFb0116200} (\cref{tab:2.1_dynCA_models}) based on the experimentally observed data are included for the comparison of the applicability of various dynamic CA models.

\begin{table*}[!htb]
  \centering
  \renewcommand\arraystretch{2.5}
    \caption{Compared dynamic contact angle models in this work.}
    \label{tab:2.1_dynCA_models}
    \begin{tabular}{ p{4cm} p{2.5cm} p{6.5cm} }
    \toprule[1.5pt]
    Molecular kinetic theory & MKT & $\theta_m = \arccos \left( \cos{\theta_0} + \frac{2 k_B T}{\sigma \lambda^2} \sinh^{-1} \left( {\frac{U}{2 \kappa^0 \lambda}} \right) \right)$ \\ 
    \midrule
    Hydrodynamic theory & HDT & $\theta_d^3 = \theta_0^3 + 9 \mathit{Ca} \beta$, with $\beta = \ln \left(\frac{L}{L_s} \right)$ \\ 
    \midrule
    \multirow{2}{*}{Combined model} & \multirow{2}{*}{Combined} & $\theta_d^3 = \theta_m^3 + 9 \mathit{Ca} \beta$, with $\beta = \ln \left(\frac{L}{L_s} \right)$, \\ 
     & & $\theta_m = \arccos \left( -\zeta \frac{U}{\sigma} + \cos{\theta_0} \right)$ \\
    \midrule
    \multirow{2}{*}{Empirical approach} & Jiang & $\frac{\cos{\theta_0}-\cos{\theta_d}}{\cos{\theta_0 +1}} = \tanh{\left(4.96Ca^{0.702}\right)}$ \\
    \cmidrule{2-3} 
     & Bracke & $\frac{\cos{\theta_0}-\cos{\theta_d}}{\cos{\theta_0 +1}} = 2 \sqrt{Ca}$ \\
    \bottomrule[1.5pt]
    \end{tabular}
\end{table*}

\section{Experiment}
\label{sec:03_experiment}

\subsection{Materials and samples}
\label{subsec:3.1_materials}
\begin{figure*}[!htb]
\centering
\def\svgwidth{0.8\textwidth}
\input{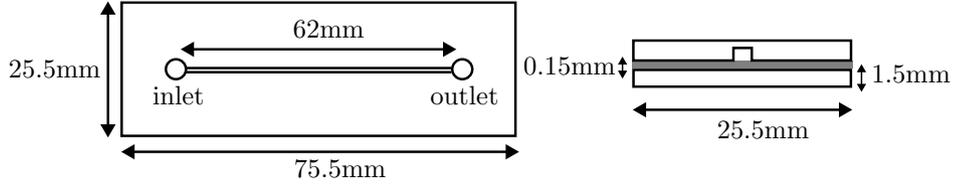}
\caption{Sketch of the test sample (straight channel). Left: top view. Right: cross section (From top to bottom: PMMA top plate with microchannel, TPU membrane and PMMA bottom plate, repectively).}
\label{fig:3.0_sketch}
\end{figure*}

\begin{figure}[!htb]
\centering
    \begin{subfigure}[t]{.48\linewidth}
        \centering
        \includegraphics[width=\linewidth]{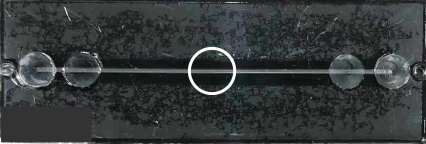}
        \caption{Straight channel}
        \label{fig:3.1.1_straight}
    \end{subfigure}
    \hfill
    \begin{subfigure}[t]{.48\linewidth}
        \centering
        \includegraphics[width=\linewidth]{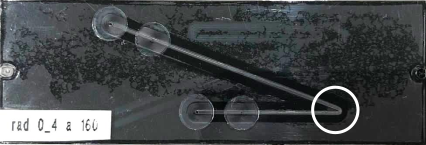}
        \caption{Variation 3}
        \label{fig:3.1.2_var3}
    \end{subfigure}
    \hfill
    \begin{subfigure}[t]{.48\linewidth}
        \centering
        \includegraphics[width=\linewidth]{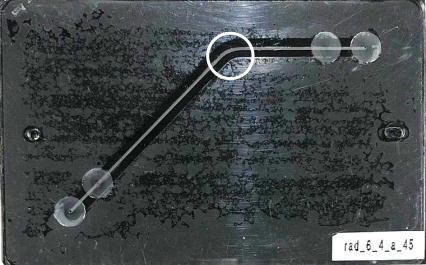}
        \caption{Variation 1}
        \label{fig:3.1.3_var1}
    \end{subfigure}
    \hfill
    \begin{subfigure}[t]{.48\linewidth}
        \centering
        \includegraphics[width=\linewidth]{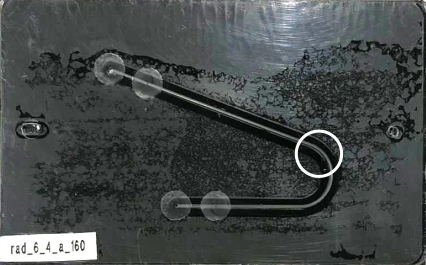}
        \caption{Variation 2}
        \label{fig:3.1.4_var2}
    \end{subfigure}
\caption{Top view of milled microchannels with different geometries and their region of interest (ROI) marked as circle.}
\label{fig:3.1_samples}
\end{figure}

For the experimental study, four test samples are employed. Each sample consists of two PolyMethyl-MethAcrylate (PMMA) microscopy slides as top and bottom plate and a thermoplastic Polyurethane (TPU) membrane. The PMMA top plate contains microchannels with a well defined geometry and dimensions and connector ports for the inlet and outlet. The geometry was manufactured by micro-milling. The bottom plate works as a support layer. As can be seen in \cref{fig:3.0_sketch}, the TPU membrane is used to seal the geometries and is firstly welded at the bottom plate and then on the top plate. All tested samples have microchannels with the same rectangular cross-section (width $\times$ depth = 0.554 mm $\times$ 0.406 mm) with microscopy shown in \cref{fig:app_microscopy}, while they can be classified as straight channel (\cref{fig:3.1.1_straight}), variation 1 (\cref{fig:3.1.3_var1}), variation 2 (\cref{fig:3.1.4_var2}) and variation 3 (\cref{fig:3.1.2_var3}) according to the geometry of the curved region. The region of interest (ROI) of all four geometries is marked as circles in \cref{fig:3.1_samples}. The angle of the curved segment $\alpha$, inside and outside radius $R_i$ and $R_o$ of the curved part is sketched in \cref{fig:3.2_curavtureRatio} and the radius values and their ratio $R_o/R_i$ are given in \cref{tab:3.1_channelGeometry}. 

\begin{minipage}[!htb]{\textwidth}
\begin{minipage}[b]{0.3\textwidth}
\centering
\begingroup%
  \makeatletter%
  \providecommand\color[2][]{%
    \errmessage{(Inkscape) Color is used for the text in Inkscape, but the package 'color.sty' is not loaded}%
    \renewcommand\color[2][]{}%
  }%
  \providecommand\transparent[1]{%
    \errmessage{(Inkscape) Transparency is used (non-zero) for the text in Inkscape, but the package 'transparent.sty' is not loaded}%
    \renewcommand\transparent[1]{}%
  }%
  \providecommand\rotatebox[2]{#2}%
  \newcommand*\fsize{\dimexpr\f@size pt\relax}%
  \newcommand*\lineheight[1]{\fontsize{\fsize}{#1\fsize}\selectfont}%
  \ifx\svgwidth\undefined%
    \setlength{\unitlength}{237.71508352bp}%
    \ifx\svgscale\undefined%
      \relax%
    \else%
      \setlength{\unitlength}{\unitlength * \real{\svgscale}}%
    \fi%
  \else%
    \setlength{\unitlength}{\svgwidth}%
  \fi%
  \global\let\svgwidth\undefined%
  \global\let\svgscale\undefined%
  \makeatother%
  \begin{picture}(1,0.59417506)%
    \lineheight{1}%
    \setlength\tabcolsep{0pt}%
    \put(0,0){\includegraphics[width=\unitlength,page=1]{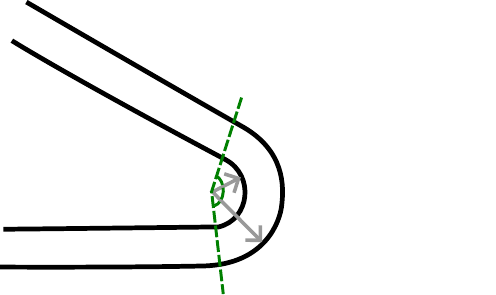}}%
    \put(0.34380078,0.1635444){\color[rgb]{0,0,0}\makebox(0,0)[lt]{\lineheight{1.25}\smash{\begin{tabular}[t]{l}$R_o$\end{tabular}}}}%
    \put(0.34889968,0.23477338){\color[rgb]{0,0,0}\makebox(0,0)[lt]{\lineheight{1.25}\smash{\begin{tabular}[t]{l}$R_i$\end{tabular}}}}%
    \put(0.47662556,0.27750905){\color[rgb]{0,0,0}\makebox(0,0)[lt]{\lineheight{1.25}\smash{\begin{tabular}[t]{l}\textcolor{teal}{$\alpha$}\end{tabular}}}}%
  \end{picture}%
\endgroup%

\captionof{figure}{Schematic illustration of the angle $\alpha$ (green), the outside radius ($R_o$) and inside radius ($R_i$) of the curved part of the example of variation 3.}
\label{fig:3.2_curavtureRatio}
\end{minipage}
\hfill
\begin{minipage}[b]{0.65\textwidth}
\centering
\begin{tabular}{c c c c c}
\toprule[1.5pt]
Microchannel & $\beta$ & $R_o$ & $R_i$ & $R_o / R_i$\\
 & (deg) & (\SI{}{mm}) &  (\SI{}{mm}) & \\
\midrule[0.5pt]
Variation 1 & 45 & 6.68 & 6.12 & 1.09\\
Variation 2 & 160 & 6.68 & 6.12 & 1.09\\
Variation 3 & 160 & 0.68 & 0.12 & 5.23\\
\bottomrule[1.5pt]
\end{tabular}
\captionof{table}{The angle of curved segment $\alpha$, outside radius $R_o$, inside radius $R_i$ of the curved part of microchannels, and their ratio $R_o / R_i$.}
\label{tab:3.1_channelGeometry}
\end{minipage}
\end{minipage}
\vspace{\belowdisplayskip}

As working fluid, de-ionized water and 50 wt\% Glycerin/water mixture are used. To enhance the visibility, Fluorescin with a molar concentration of $\SI{5e-5}{mol/L}$ is added, ensuring not to influence the fluid physical properties. The density, the dynamic viscosity and the surface tension were internally determined and are given in \cref{tab:3.2_physicalProperties}.  It should be mentioned that the 50 wt\% Glycerin/water mixture is applied only on the straight channel (\cref{fig:3.1.1_straight}), while water is applied for all three curved channels due to its relevance to the industrial applications.

\begin{table*}[h]
\centering
\caption{Physical properties of the working fluids and surface tension with air.}
\label{tab:3.2_physicalProperties}
\begin{tabular}{c c c c}
\toprule[1.5pt]
Working fluid & density & dyn. viscosity & surface tension  \\
@ 22°C & $\rho\ (\SI{}{kg/m^3})$ & $\mu\ (\SI{}{Pa \cdot s})$ & $\sigma\ (\SI{}{N/m})$ \\
\midrule[0.5pt]
Water & 998.03 & 1.000e-3 & 72.74e-3 \\
50 wt\% Glycerin/water mixture & 1142 & 8.026e-3 & 68.12e-3 \\
\bottomrule[1.5pt]
\end{tabular}
\end{table*}

\subsection{Experimental setup}
\label{subsec:3.2_setup}

\begin{figure*}[!htb]
\centering
\def\svgwidth{0.8\textwidth}
\begingroup%
  \makeatletter%
  \providecommand\color[2][]{%
    \errmessage{(Inkscape) Color is used for the text in Inkscape, but the package 'color.sty' is not loaded}%
    \renewcommand\color[2][]{}%
  }%
  \providecommand\transparent[1]{%
    \errmessage{(Inkscape) Transparency is used (non-zero) for the text in Inkscape, but the package 'transparent.sty' is not loaded}%
    \renewcommand\transparent[1]{}%
  }%
  \providecommand\rotatebox[2]{#2}%
  \newcommand*\fsize{\dimexpr\f@size pt\relax}%
  \newcommand*\lineheight[1]{\fontsize{\fsize}{#1\fsize}\selectfont}%
  \ifx\svgwidth\undefined%
    \setlength{\unitlength}{567.73194551bp}%
    \ifx\svgscale\undefined%
      \relax%
    \else%
      \setlength{\unitlength}{\unitlength * \real{\svgscale}}%
    \fi%
  \else%
    \setlength{\unitlength}{\svgwidth}%
  \fi%
  \global\let\svgwidth\undefined%
  \global\let\svgscale\undefined%
  \makeatother%
  \begin{picture}(1,0.6625374)%
    \lineheight{1}%
    \setlength\tabcolsep{0pt}%
    \put(0,0){\includegraphics[width=\unitlength,page=1]{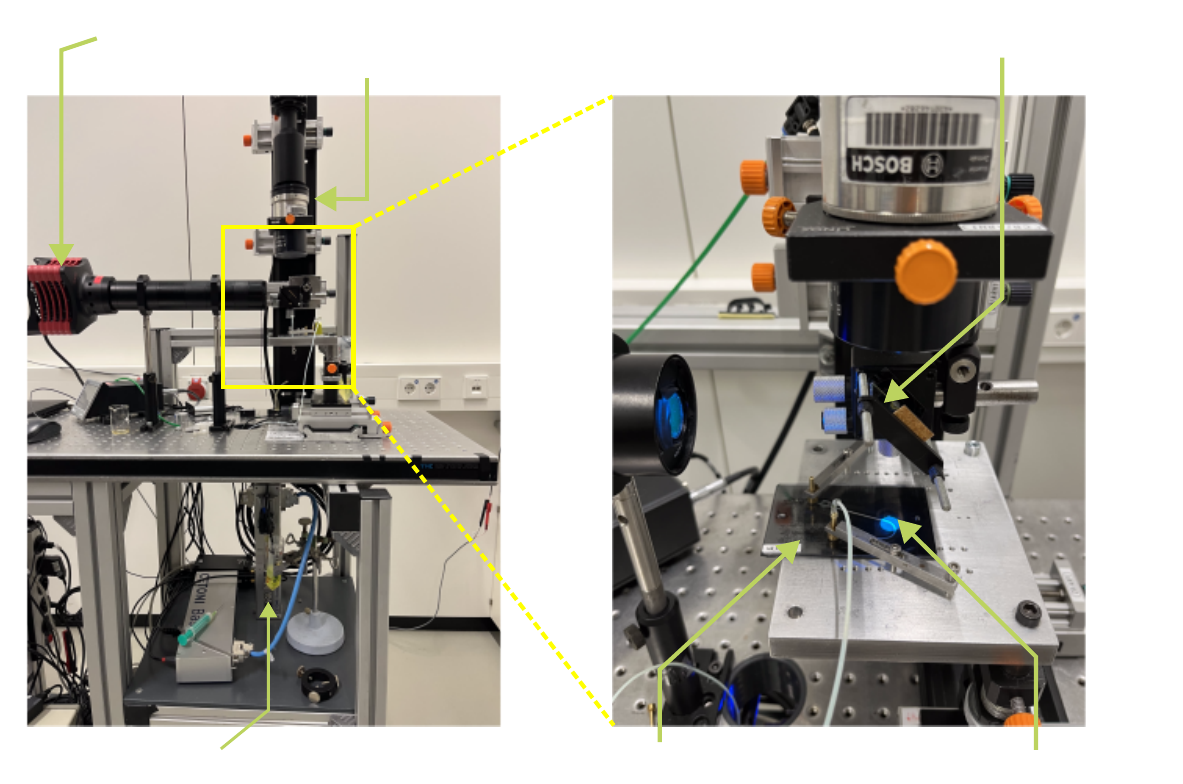}}%
    \put(0.02103067,0.01907408){\color[rgb]{0,0,0}\makebox(0,0)[lt]{\lineheight{1.25}\smash{\begin{tabular}[t]{l}Syringe pump\end{tabular}}}}%
    \put(-0.0014872,0.64439129){\color[rgb]{0,0,0}\makebox(0,0)[lt]{\lineheight{1.25}\smash{\begin{tabular}[t]{l}High power LED\end{tabular}}}}%
    \put(0.23106262,0.64238116){\color[rgb]{0,0,0}\makebox(0,0)[lt]{\lineheight{1.25}\smash{\begin{tabular}[t]{l}High-speed camera\\with objectiv\end{tabular}}}}%
    \put(0.75698658,0.62104849){\color[rgb]{0,0,0}\makebox(0,0)[lt]{\lineheight{1.25}\smash{\begin{tabular}[t]{l}Dichroic mirror\end{tabular}}}}%
    \put(0.43083139,0.01145312){\color[rgb]{0,0,0}\makebox(0,0)[lt]{\lineheight{1.25}\smash{\begin{tabular}[t]{l}Test sample (var2)\end{tabular}}}}%
    \put(0.72505555,0.00547838){\color[rgb]{0,0,0}\makebox(0,0)[lt]{\lineheight{1.25}\smash{\begin{tabular}[t]{l}Region of interest (ROI)\end{tabular}}}}%
  \end{picture}%
\endgroup%

\caption{Experimental setup. Left: test bench. Right: zoom-in view with region of interest (ROI).}
\label{fig:3.3_setup}
\end{figure*}

Figure \ref{fig:3.3_setup} shows the experimental setup of the whole test bench on the left side and a zoom-in view on the right side. The liquid flow is controlled by a syringe pump (CETONI Nemesys S). To clearly visualize the interface motion a FITC-Fluorescence setup in combination with a high-speed camera (NX4-S3, LDT) is used. Therefore, a high power LED is chosen (SOLIC-3C) as light source. To strictly define the excitation light spectrum an excitation filter is used (483x31\SI{}{nm}). The filtered light beam is reflected on a dichroic mirror (refl.: 470 - 490\SI{}{nm}, trans: 505 - 800\SI{}{nm}) and excites the FITC molecules in the working fluid. The illuminated areas at the sample is marked with a circle in \cref{fig:3.1_samples}. The phase-shifted emitted light is then transmitted through the dichroic mirror to the camera again. To remove possible effects by ambient light and reflections due to the emission beam an emission filter is included in the camera tubing system with a transmission spectrum of 530 $\times$ 45\SI{}{nm}. 

The ROI is visualized by 1024x1024 pixels with a resolution of $\Delta x = \SI{6.566e-6}{m}$. The frame rate was chosen in dependency of the operation point and the flow rate (\cref{tab:3.3_opticalParameter}). The image grabbing is accomplished with the software MotionStudio.

\begin{table*}[!htb]
\centering
\caption{Investigated volumetric flow rates, flow velocities in experiments and the corresponding optical frame rates.}
\label{tab:3.3_opticalParameter}
\begin{tabular}{c c c}
\toprule[1.5pt]
Volumetric flow rate & Flow velocity & Frame rate \\
$\dot{V} \ (\SI{e-9}{m^3/s})$ & $U \ (\SI{e-3}{m/s})$ & (fps)\\
\midrule[0.5pt]
0.025, 0.11 & 0.11, 0.49 & 30 \\
0.25, 0.67 & 1.11, 2.98 & 60 \\
1.1, 1.8 & 4.89, 8.00 & 100 \\
2.5 & 11.11 & 200 \\
2.7, 3.1, 3.6, 4, 4.5 & 12.22, 13.78, 16.00, 17.78, 20.00 & 300 \\
\bottomrule[1.5pt]
\end{tabular}
\end{table*}

\subsection{Experimental procedure}
\label{subsec:3.3_procedure}
For all operation points a constant flow rate was chosen as listed in \cref{tab:3.3_opticalParameter}. The liquid enters the channel at the inlet port and leaves the system via the outlet port (\cref{fig:3.0_sketch}). The image recording was manually started. The determination of the dynamic CA was done afterwards using the post-processing procedure, as described in \cref{sec:04_automatedAnalysis}. After each test, the channel is cleaned with de-ionized water flow and then dried with air flow for 5 minutes. Each test is repeated three times to ensure the reproducibility and reliability of the experimental findings, confirmed by very low variance in results from repeated experiments.

To further characterize the liquid-solid-interaction the static CA was determined for every specific sample in addition. Therefore, the working fluid was pumped into the channel with the smallest volumetric flow rate $\dot{V} = \SI{0.025e-9}{m^3/s}$ up to the middle of the ROI. Then, the pump was stopped and after a waiting time of three minutes the images were recorded for the static CA measurements. It should be noted due to the optical effect of sharp edges of the rectangular channel cross section and the limitation of 2D top-down visualization of the 3D fluid interface, the static CA values measured in channel can deviate from the values measured with a fluid droplet on a solid substrate. Nevertheless, the rivulet effect along sharp corner in rectangular channel is not relevant here as explained in \cref{appdendice1}.

\section{Automated dynamic contact angle analysis}
\label{sec:04_automatedAnalysis}
In this section, the proposed automated dynamic contact angle analysis is explained in four steps: interface detection (\cref{subsec:4.2_interfaceDetection}), contact angle measurement (\cref{subsec:4.3_contactAngle}), outlier removal (\cref{subsec:4.4_outlierRemoval}) and output (\cref{subsec:4.5_output}). The developed image processing procedure provides a technique for automated and robust measurement of the interfacial characteristics through microchannels with arbitrary geometries. This method is particularly advantageous for experiments with non-transparent samples, where extracting interface characteristics poses a challenge. The source code version used in this manuscript is archived in a jupyter notebook on GitHub \citep{automatedImageAnalysis}.

\begin{figure*}[!htb]
\tikzstyle{input} = [rectangle, rounded corners, 
minimum width=3cm, minimum height=1cm,
text centered, text width=4cm,
draw=black, fill=gray!15]

\tikzstyle{interface} = [rectangle, rounded corners, 
minimum width=3cm, minimum height=1cm,
text centered, text width=6cm,
draw=black, fill=green!15]

\tikzstyle{dynCA} = [rectangle, rounded corners, 
minimum width=3cm, minimum height=1cm,
text centered, text width=6cm,
draw=black, fill=blue!15]

\tikzstyle{outlier} = [rectangle, rounded corners, 
minimum width=3cm, minimum height=1cm,
text centered, text width=5cm,
draw=black, fill=red!15]

\tikzstyle{output} = [rectangle, rounded corners, 
minimum width=3cm, minimum height=1cm,
text centered, text width=5cm,
draw=black, fill=yellow!15]

\tikzstyle{output2} = [rectangle, rounded corners, 
minimum width=3cm, minimum height=1cm,
text centered, text width=3.3cm,
draw=black, fill=yellow!15]

\begin{tikzpicture}[node distance=2cm, auto]
    \node (input1)[input]{Pre-processed images};
    \node (interface1)[interface, below of=input1, node distance=1.5cm]{Extract the complete contours of images at $t_n$ and $t_{n+1}$ as $c_n$ and $c_{n+1}$};
    \node (interface2)[interface, below of=interface1, node distance=2cm]{Calculate the contour difference $c_d = c_{n+1} - c_n$ with the momentum center of $c_d$ as point $C (C_x, C_y)$ with $C_x = M_{10}/M_{00}$ and $C_y = M_{01}/M_{00}$};
    \node (interface3)[interface, below of=interface2, node distance=2cm]{Find the closest point $P$ to the point $C$ on the contour $c_n$};
    \node (interface4)[interface, below of=interface3, node distance=2cm]{From the point $P$ search for the first two ``overlapping" points of the contour $c_n$ and $c_{n+1}$ as the points $a_n$ and $b_n$};
    \node (interface5)[interface, below of=interface4, node distance=2.2cm]{The interface of the contour $c_n$ can be detected with the two boundary points $a_n$ and $b_n$};
    \node (dynCA1)[dynCA, right of=input1, xshift=7cm]{From the boundary points $a_n$ and $b_n$ choose the closest N pixels on the contour $c_n$ as tangents};
    \node (dynCA2)[dynCA, below of=dynCA1, node distance=2cm]{Calculate the angles between the tangents as the contact angle between the interface and the wall};
    \node (outlier1)[outlier, below of=dynCA2, node distance=1.8cm]{Channel-width constraint};
    \node (outlier2)[outlier, below of=outlier1, node distance=1.7cm]{Contact-angle constraint};
    \node (output1)[output, below of=outlier2, node distance=1.8cm]{Save the processed images with the contact angle information};
    \node (node1)[coordinate, below of=output1, node distance=1cm]{};
    \node (output2)[output2, below left=0.6cm and 0.3cm of node1, node distance=3.7cm]{Calculate the interface movement using image resolution and frame rate, save the results in csv file};
    \node (output3)[output2, below right=0.6cm and 0.3cm of node1, node distance=3.7cm]{Plot the temporal evolution of the interface movement and dynamic contact angle};
    \draw [->] (input1) -- (interface1);
    \draw [->] (interface1) -- (interface2);
    \draw [->] (interface2) -- (interface3);
    \draw [->] (interface3) -- (interface4);
    \draw [->] (interface4) -- (interface5);
    \draw [->] (interface5) -| ++(4,4) |- (dynCA1);
    \draw [->] (dynCA1) -- (dynCA2);
    \draw [->] (dynCA2) -- (outlier1);
    \draw [->] (outlier1) -- (outlier2);
    \draw [->] (outlier2) -- (output1);
    \draw [-] (output1) -- (node1);
    \draw [->] (node1.south) |-++ (0,0) -| (output2);
    \draw [->] (node1.south) |-++ (0,0) -| (output3);
\end{tikzpicture}
\caption{Flow chart of automated image processing procedure: input (gray), interface detection (green), contact angle measurement (blue), outlier removal (red) and output (yellow).}
\label{fig:4.1_flowChart}
\end{figure*}
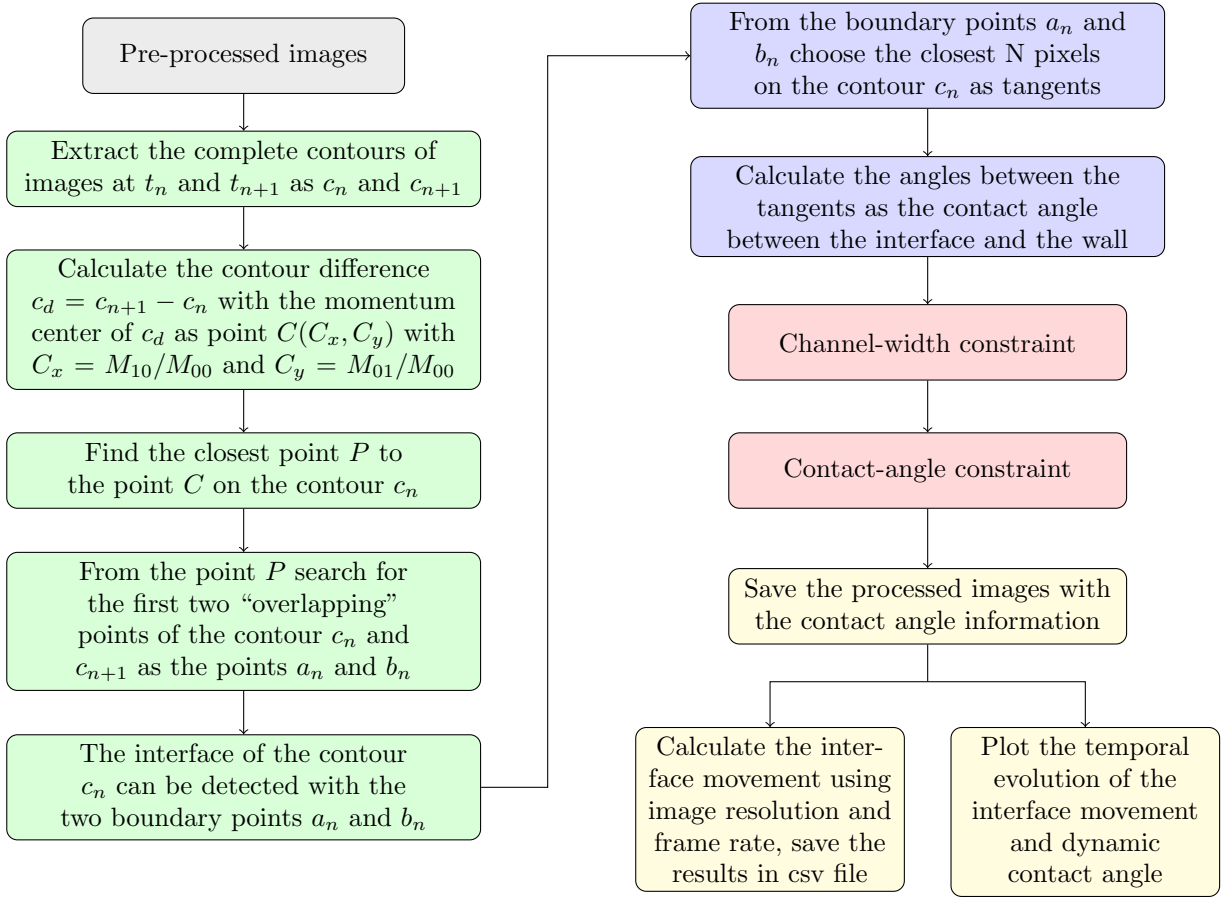

\subsection{Interface detection}
\label{subsec:4.2_interfaceDetection}
\begin{figure*}[!htb]
\centering
    \begin{subfigure}[t]{.4\linewidth}
        \centering
        \includegraphics[width=\linewidth]{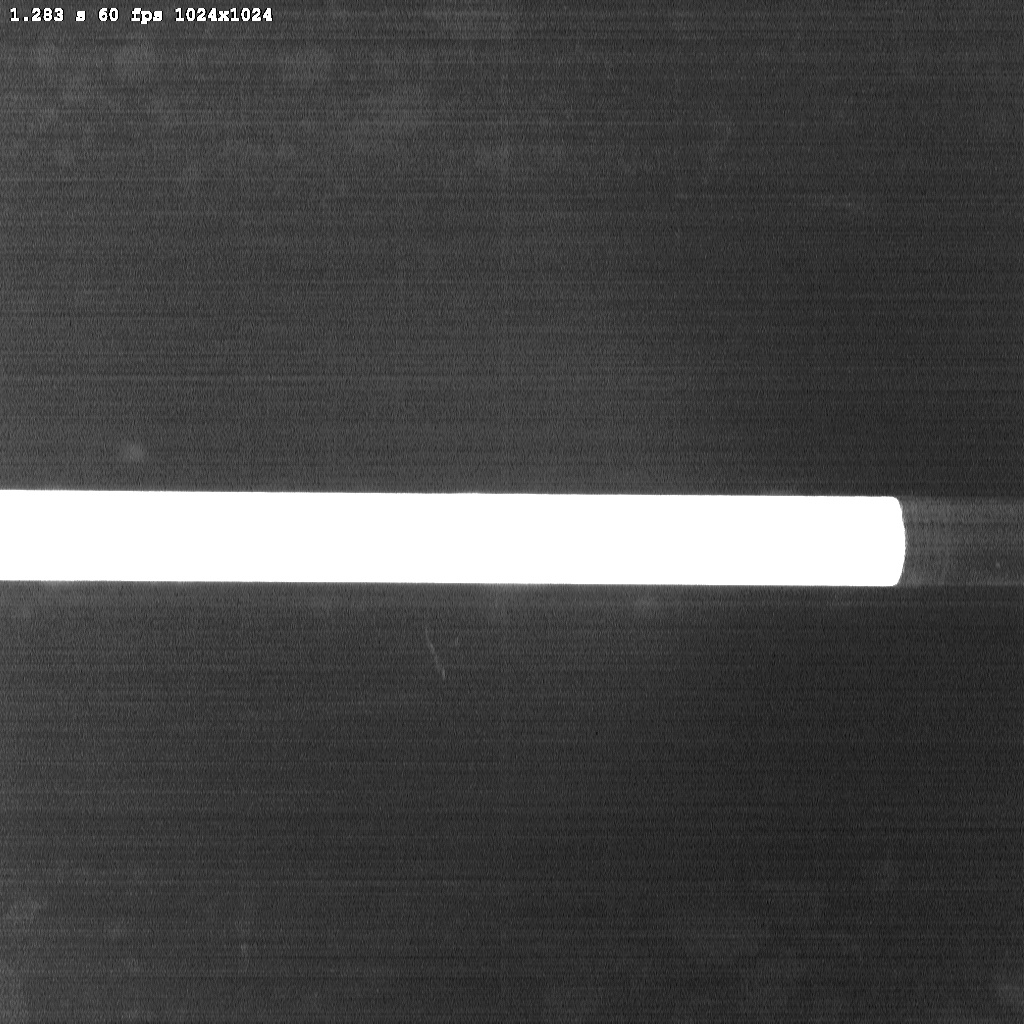}
        \caption{Pre-processed image at $t_n$.}
        \label{fig:4.2.1_preprocessN}
    \end{subfigure}
    \hfill
    \begin{subfigure}[t]{.4\linewidth}
    \centering
    \includegraphics[width=\linewidth]{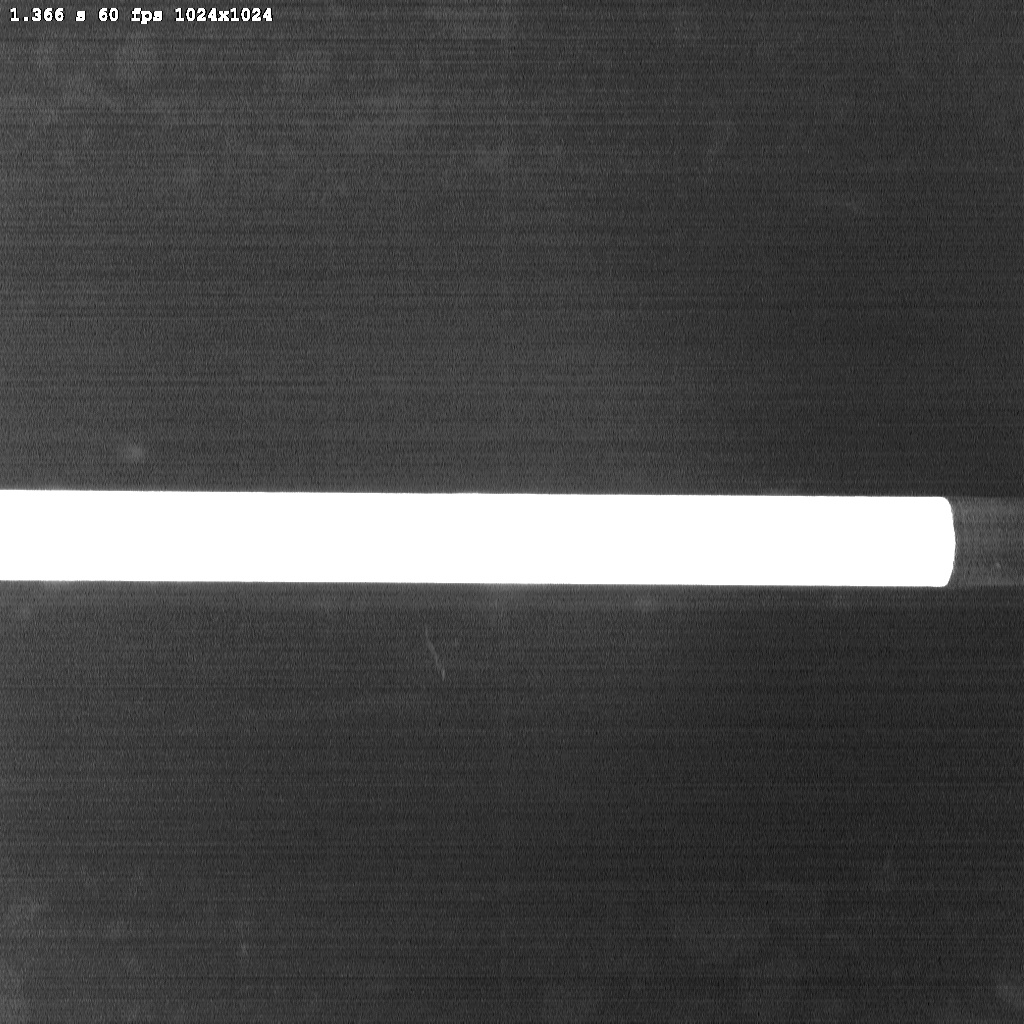}
    \caption{Pre-processed image at $t_{n+1}$.}
    \label{fig:4.2.2_preprocessN+1}
    \end{subfigure}
    \hfill
    \begin{subfigure}[t]{.31\linewidth}
        \centering
        \includegraphics[width=\linewidth]{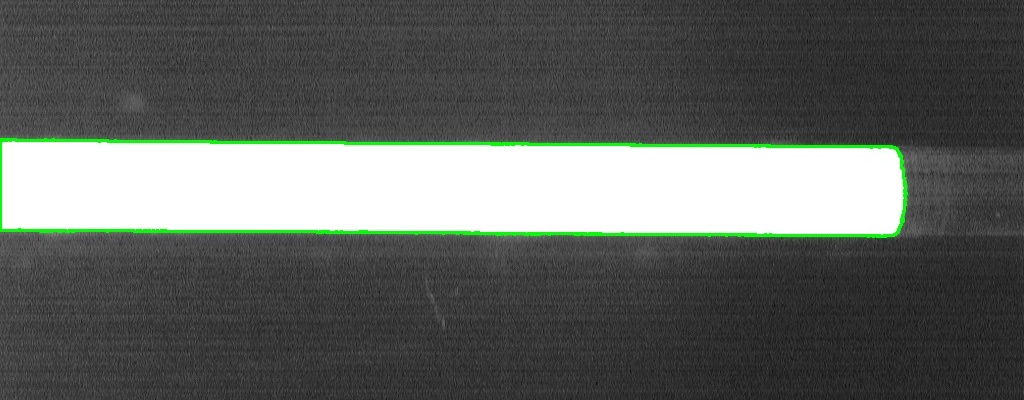}
        \caption{Complete contour of the image at $t_n$ as $c_n$.}
        \label{fig:4.2.3_contourN}
    \end{subfigure}
    \hfill
    \begin{subfigure}[t]{.31\linewidth}
        \centering
        \includegraphics[width=\linewidth]{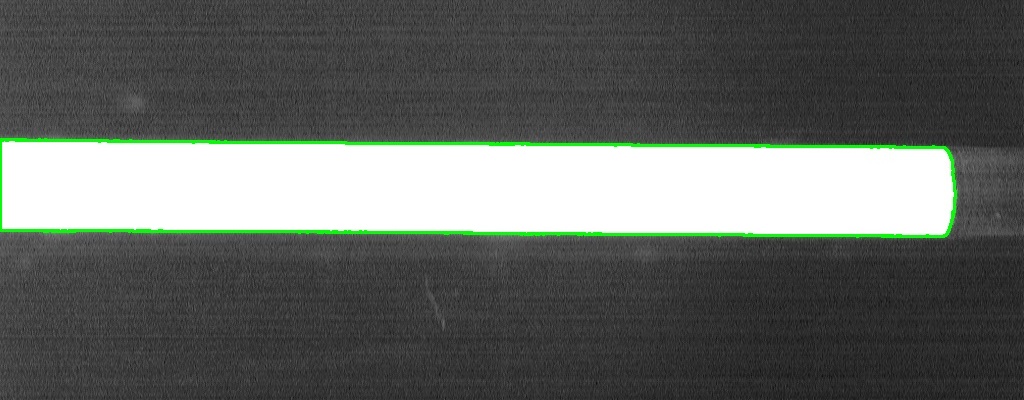}
        \caption{Complete contour of the image at $t_{n+1}$ as $c_{n+1}$.}
        \label{fig:4.2.3_contourN+1}
    \end{subfigure}
    \hfill
    \begin{subfigure}[t]{.31\linewidth}
        \centering
        \includegraphics[width=\linewidth]{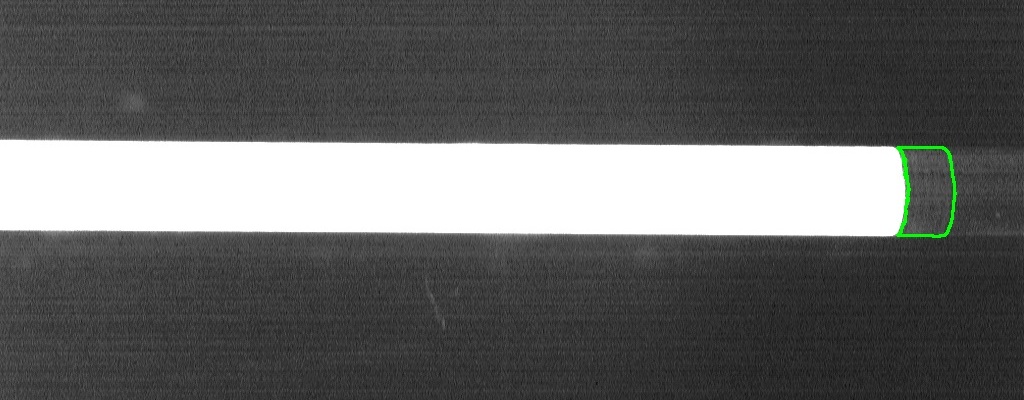}
        \caption{Contour difference $c_d$ between $c_n$ and $c_{n+1}$.}
        \label{fig:4.2.4_contourdiff}
    \end{subfigure}
    \hfill
    \begin{subfigure}[t]{.49\linewidth}
        \centering
        \includegraphics[width=\linewidth]{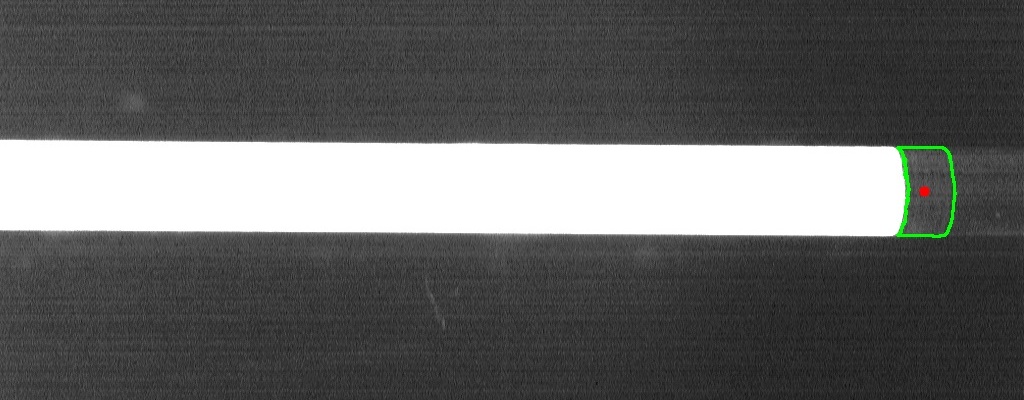}
        \caption{The momentum center $C$ (red point) of the contour difference $c_d$.}
        \label{fig:4.2.5_centroid}
    \end{subfigure}
    \hfill
    \begin{subfigure}[t]{.49\linewidth}
        \centering
        \includegraphics[width=\linewidth]{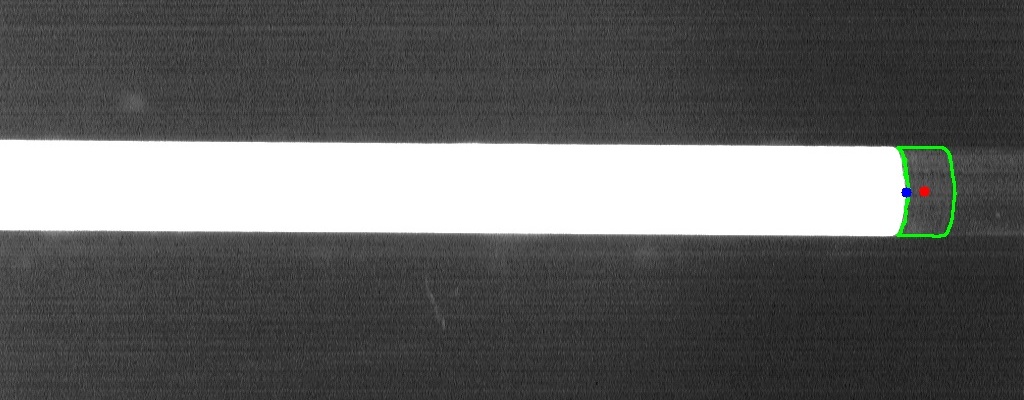}
        \caption{The start point $P$ of the following iterations (blue point).}
        \label{fig:2.4.5_pointP}
    \end{subfigure}
    \hfill
    \begin{subfigure}[t]{.49\linewidth}
        \centering
        \includegraphics[width=\linewidth]{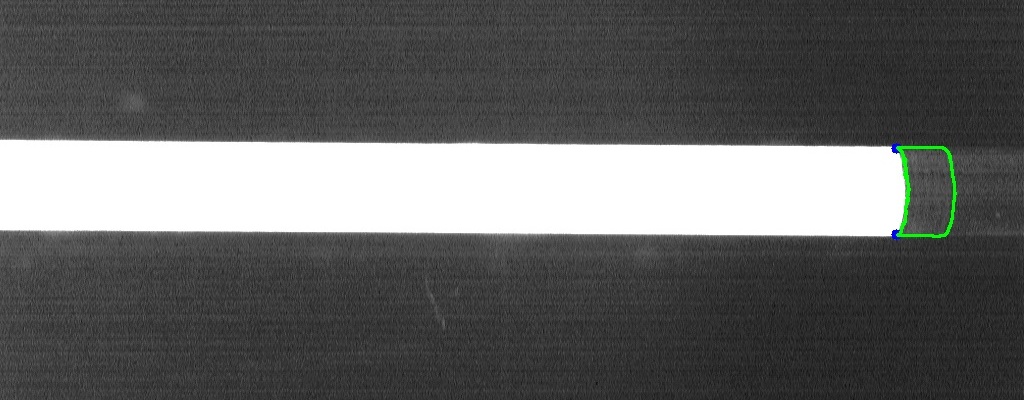}
        \caption{The boundary points of the interface $a_n$ (blue point on top) and $b_n$ (blue point on bottom).}
        \label{fig:2.4.6_boundaryPoints}
    \end{subfigure}
    \hfill
    \begin{subfigure}[t]{.49\linewidth}
        \centering
        \includegraphics[width=\linewidth]{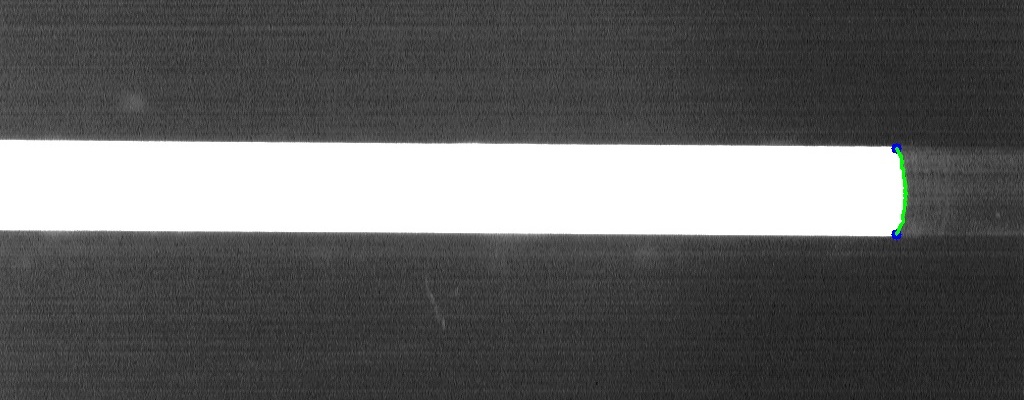}
        \caption{The interface of the image at $t_n$.}
        \label{fig:2.4.7_interface}
    \end{subfigure}
\caption{The interface detection procedure.}
\label{fig:4.2_interfaceDetection}
\end{figure*}

To enable the contact angle measurements between interface and walls, the fluid interface need to be firstly identified and extracted. Figure \ref{fig:4.1_flowChart} illustrates the sequential workflow of the automated image processing used in this work. Figure \ref{fig:4.2_interfaceDetection} provides an visual overview of the interface detection procedure with two successive images as examples. To optimize space utilization, only the first two pre-processed images are presented in their original size, while the subsequent images have been cropped for conciseness. 

The complete contours of the fluid flow of two successive images at $t_n$ and $t_{n+1}$ are extracted as $c_n$ (\cref{fig:4.2.3_contourN}) and $c_{n+1}$ (\cref{fig:4.2.3_contourN+1}). A contour in this context is a list of index pairs $(i,j)$ that specify the pixels on the image that belong to the contour. Then, a set difference $c_d$ between contours $c_n$ and $c_n+1$ is calculated as $c_d = c_{n+1} - c_n$ (\cref{fig:4.2.4_contourdiff}) with the momentum center as point $C (C_x, C_y)$ with $C_x = M_{10}/M_{00}$ and $C_y = M_{01}/M_{00}$ as shown in \cref{fig:4.2.5_centroid}. Here, $M = \sum_x \sum_y x^i y^j I(x,y)$ denotes the momentum of the contour difference $c_d$ with the $I(x,y)$ for the intensity of each pixel. To enable the subsequent iterations for interface detection, the closest point on the contour $c_n$ to the momentum center $C$ is found as the start point and is drawn in \cref{fig:2.4.5_pointP} as the point $P$. From the point $P$, the iteration starts to search for the first two "overlapping" points $a_n$ and $b_n$ (\cref{fig:2.4.6_boundaryPoints}) of contour $c_n$ and $c_{n+1}$, which should be the boundary points of the interface of contour $c_n$. It should be emphasized that the  ``overlapping" here signifies the distance between two points smaller than a threshold value, which is given as a predefined parameter by the user. With the detected boundary points $a_n$ and $b_n$, the interface contour of image at $t_n$ can be obtained as shown in \cref{fig:2.4.7_interface}.

\subsection{Contact angle measurement}
\label{subsec:4.3_contactAngle}
\begin{figure*}[!htb]
\centering
    \begin{subfigure}[t]{.48\linewidth}
        \centering
        \includegraphics[width=\linewidth]{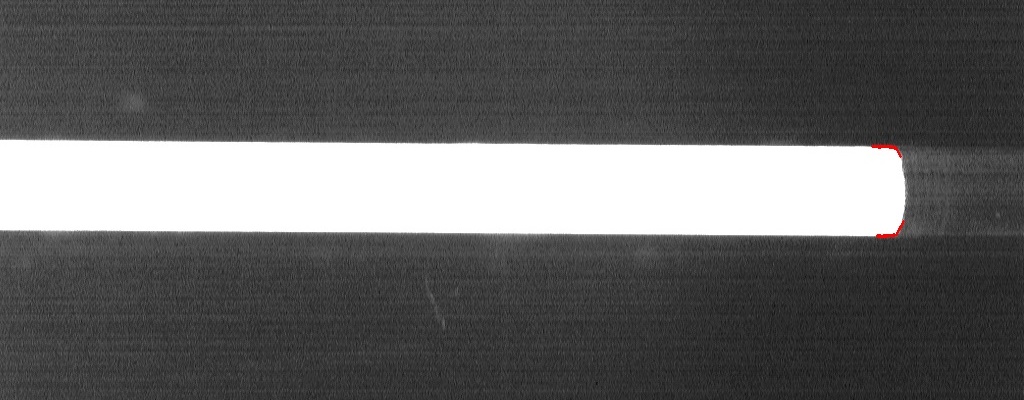}
        \caption{Local tangents at boundary points $a_n$ and $b_n$ with the closest 10 pixels.}
        \label{fig:4.3.1_tangents}
    \end{subfigure}
    \hfill
    \begin{subfigure}[t]{.48\linewidth}
        \centering
        \includegraphics[width=\linewidth]{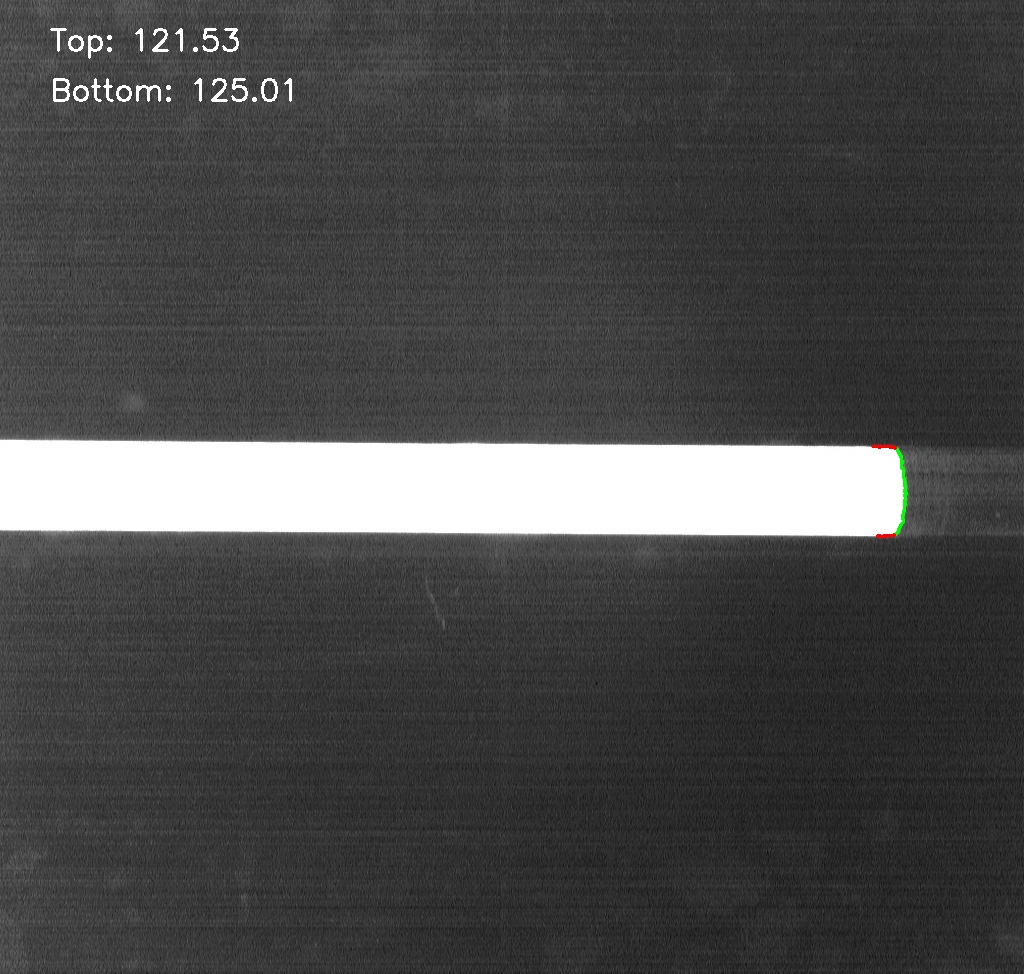}
        \caption{Calculated contact angles displayed on the image.}
        \label{fig:4.3.2_contactAngles}
    \end{subfigure}
\caption{Dynamic contact angle measurement procedure.}
\label{fig:4.3_CAmeasurement}
\end{figure*}

To measure the contact angle, the local tangents at boundary points $a_n$ and $b_n$ are identified in this proposed method. To derive these tangents, the algorithm selects the closest N pixels along the contour $c_n$ originating from the boundary points $a_n$ and $b_n$ (\cref{fig:4.3.1_tangents}). The contact angle is subsequently calculated as the angle between the two selected tangents as presented in \cref{fig:4.3.2_contactAngles}.

\subsection{Outlier removal}
\label{subsec:4.4_outlierRemoval}
Due to a small variance of the image qualities, the outlier removal is provided with two options in an intuitive way: channel-width constraint and contact-angle constraint. This step provides flexibility in addressing variations in data quality and ensures a more robust analysis.

\paragraph{Channel-width constraint}
The distance between the boundary points $a_n$ and $b_n$ should be equal to the microchannel width, which can be measured using microscope or camera precisely. Therefore, the microchannel width with a user pre-defined tolerance can be used as a criterion for the outlier removal.

\paragraph{Contact-angle constraint}
Knowing the static contact angle by previous experiments and the volumetric flow rate range allows the usage of a user defined reasonable contact angle range as the outlier removal criterion.

\subsection{Output}
\label{subsec:4.5_output}
The processed images with the detected interface and the corresponding contact angle information are saved in a new directory with the same file name as the original images. The interface movement is calculated using the image resolution and frame rate information, and then saved with the contact angle together in a CSV file. In addition, the temporal evolution of the meniscus displacement and dynamic contact angle can be visualized as plots.

To utilize the procedure for further analysis, the algorithm is firstly validated in \cref{subsec:5.1_validation}. Its automation through Python not only enhances efficiency, significantly reducing processing time, but also facilitates direct comparisons to manually measured data, thereby ensuring the method's validity and reliability. 

In spite of the fact that three parameters for the procedure should be provided by the user: \texttt{thres\_value} as the threshold value for extracting the contour, \texttt{thres\_dist} as a distance threshold for defining the boundary points $a_n$ and $b_n$ and \texttt{n\_pixels} as the number of pixels used to fit the local tangent of interface, they are confirmed to be less sensible to channel geometries. For example, in this study \texttt{thres\_value} with a value of 160 and \texttt{n\_pixels} with a value of 10 for local tangents are kept constant for all images, while the \texttt{thres\_dist} varies from 1 to 2.5 with an interval of 0.5. Nevertheless, an inappropriate range for the parameters will affect the accuracy of this method. More specifically, \texttt{thres\_value} is associated with correct and complete contour detection, while different values of \texttt{n\_pixels} might lead to a distinct discrepancy between ``local" and ``global" CA measurements. The \texttt{n\_pixels} = 10 in this work is chosen to have a comparable perspective with the manual measurements and is determined for future experiments using manual validation with a very small sub-set of images. Furthermore, the relatively more sensible \texttt{thres\_dist} is responsible for the accurate detection of boundary points, meaning a larger \texttt{thres\_dist} can cause the iteration process to stop earlier with the boundary points are still far away from the channel walls then an incomplete interface, while an excessive interface with walls included might stem from the smaller \texttt{thres\_dist}. We found that validation and calibration of the automatic CA detection procedure can be done efficiently for a small sub-set of images, and used automatically to detect dynamic CA in an entire wetting process, on a very large image sequence. In addition, this method might not be suitable for images with either substantial or minimal differences. For example, images taken with extremely low frame rates in very curved microchannels might lead to too-large and non-convex contour differences, making the search for the nearest point of the previous contour impossible. Another extreme are too high frame rates, that would result to practically overlapping contours and near-zero differences between two subsequent contours. As for the algorithm parameters, the frame rate is checked and adjusted during the validation and calibration process with a small sequence of images.

\section{Results and discussion}
\label{sec:05_results}

\subsection{Validation of automated analysis}
\label{subsec:5.1_validation}

\begin{figure}[!htb]
    \centering
    \includegraphics[width=\linewidth]{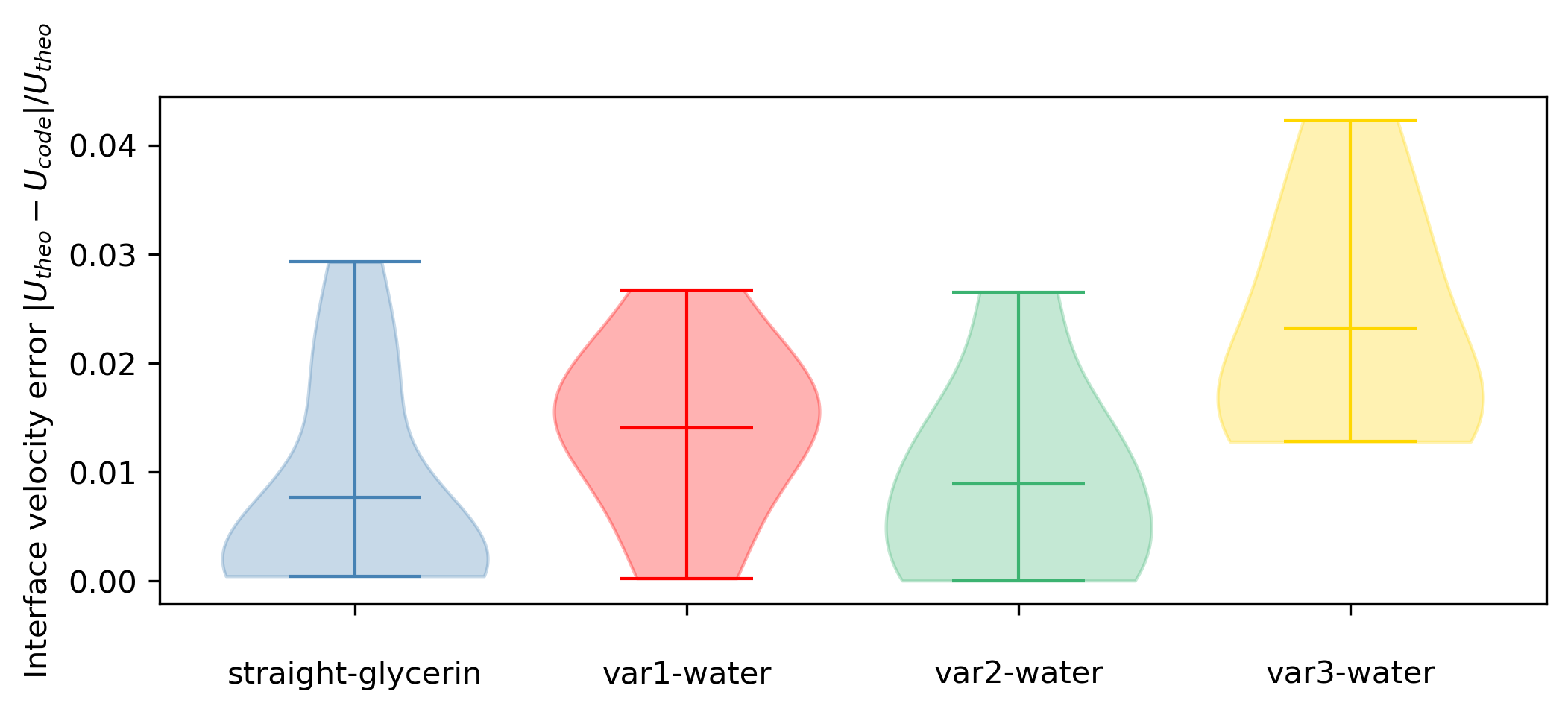}
    \caption{Comparison of theoretical interface velocity $U_{theo}$ and calculated interface velocity from measurements $U_{code}$ for interface detection validation as violin plots. Top line: maximum. Middle line: mean. Bottom line: minimum. }
    \label{fig:5.1_vali_velocity}
\end{figure}

\paragraph{Interface detection}
A correct interface detection procedure (\cref{subsec:4.2_interfaceDetection}) is fundamental for the accuracy of the automated dynamic contact angle (CA) analysis. The interface velocity is selected to be the validation parameter as a consequence of the volumetric flow rate controlled flow by the syringe pump. The theoretical interface velocity $U_{theo} = \dot{V}/A$ can be calculated using the cross section dimension of microchannels measured using a microscope. As mentioned in \cref{subsec:4.5_output}, the temporal evolution of the meniscus displacement (top and bottom) is obtained as output, of which the mean value can be further analyzed for the interface velocity $U_{code}=(x_n-x_0)/(t_n-t_0)$. The comparison between the theoretical and measured interface velocity in the form of velocity error $|U_{theo}-U_{code}|/U_{theo}$ of all tested flow rates is summarized as violin plots in \cref{fig:5.1_vali_velocity}. It is clearly visible that the algorithm captures the moving interface of serial images precisely with the maximum velocity error below 5\% and the mean velocity error smaller than 3\%.

\begin{figure}[!htb]
\centering
\includegraphics[width=\linewidth]{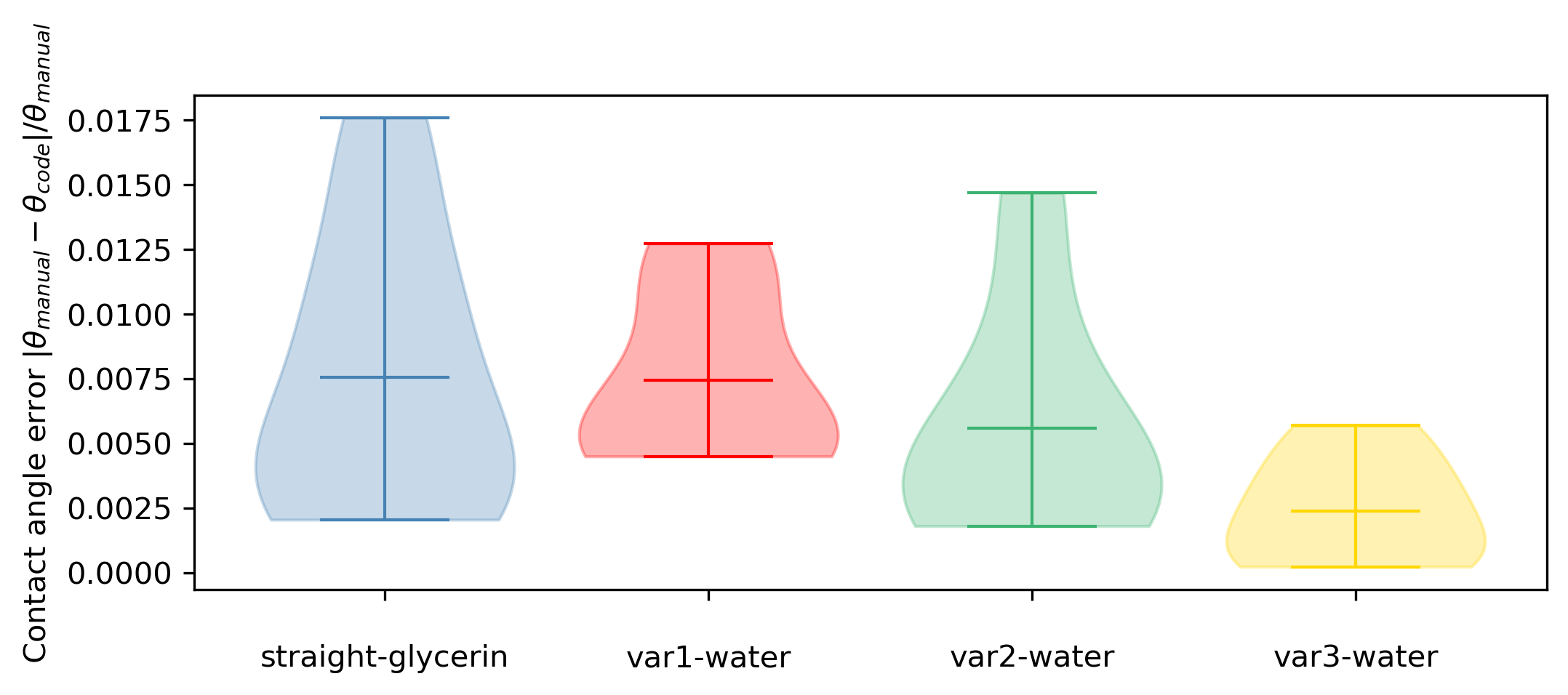}
\caption{The relative dynamic contact angle error comparing manual measurements $\theta_{manual}$ and automated measurements $\theta_{code}$ as violin plots with case-variant distribution. Top line: maximum. Middle line: mean. Bottom line: minimum.}
\label{fig:5.2_vali_theta}
\end{figure}

\paragraph{Contact angle measurement}
The automated dynamic CA measurement via local tangents is validated here by comparison with the manually measured results. To ensure the reproducibility and reliability of the manual data, three random images of each test are chosen and the CAs at both channel sides of each image are measured three times with \cite{imageJ}, respectively, of which the average value is then picked to be the manually measured as dynamic CA $\theta_{manual}$. As explained in \cref{subsec:4.3_contactAngle}, the number of the closest pixels for the local tangents is chosen to be $N=10$ for the dynamic CA calculation $\theta_{code}$. Figure \ref{fig:5.2_vali_theta} summarizes the results of all considered flow rates (\cref{tab:3.3_opticalParameter}) as violin plots. The violin plots reveal that the results from the algorithm are quite comparable to the manual measurements and the relative CA measurement error 
\begin{equation}
	e_{\theta_A} = \frac{|\theta_{manual}-\theta_{code}|}{\theta_{manual}}
\end{equation}
for all tests is below 2\%.

Based on the above validation, all experimental results shown in the following sections refer to the data from the automated analysis.

\subsection{Dynamic contact angle model}
\label{subsec:5.2_dynCAmodel}

\begin{figure*}[!htb]
\centering
    \begin{subfigure}[t]{.44\linewidth}
        \centering
        \includegraphics[width=\linewidth]{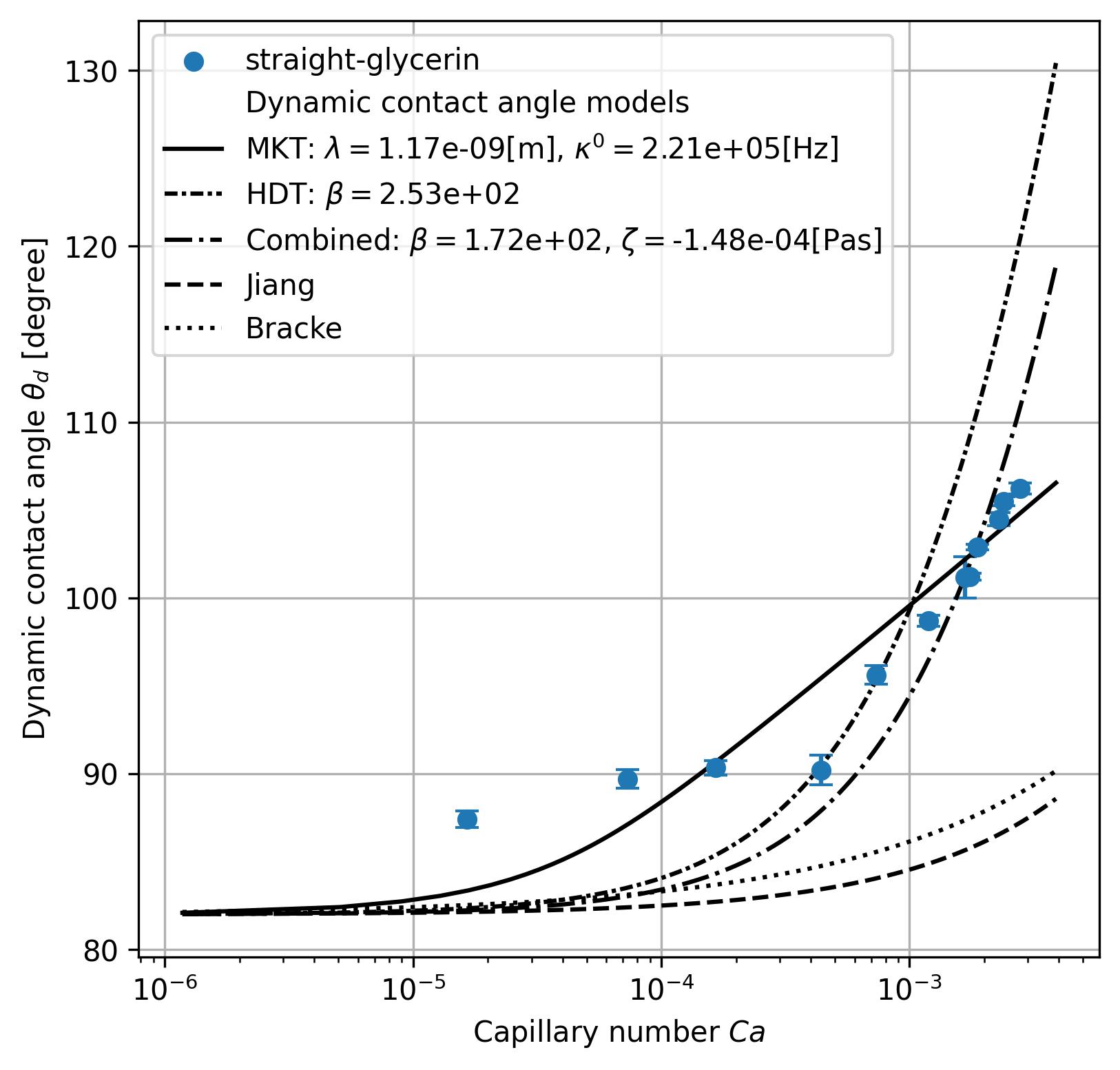}
        \caption{Straight channel}
        \label{fig:5.3.1_straight}
    \end{subfigure}
    \hfill
    \begin{subfigure}[t]{.44\linewidth}
        \centering
        \includegraphics[width=\linewidth]{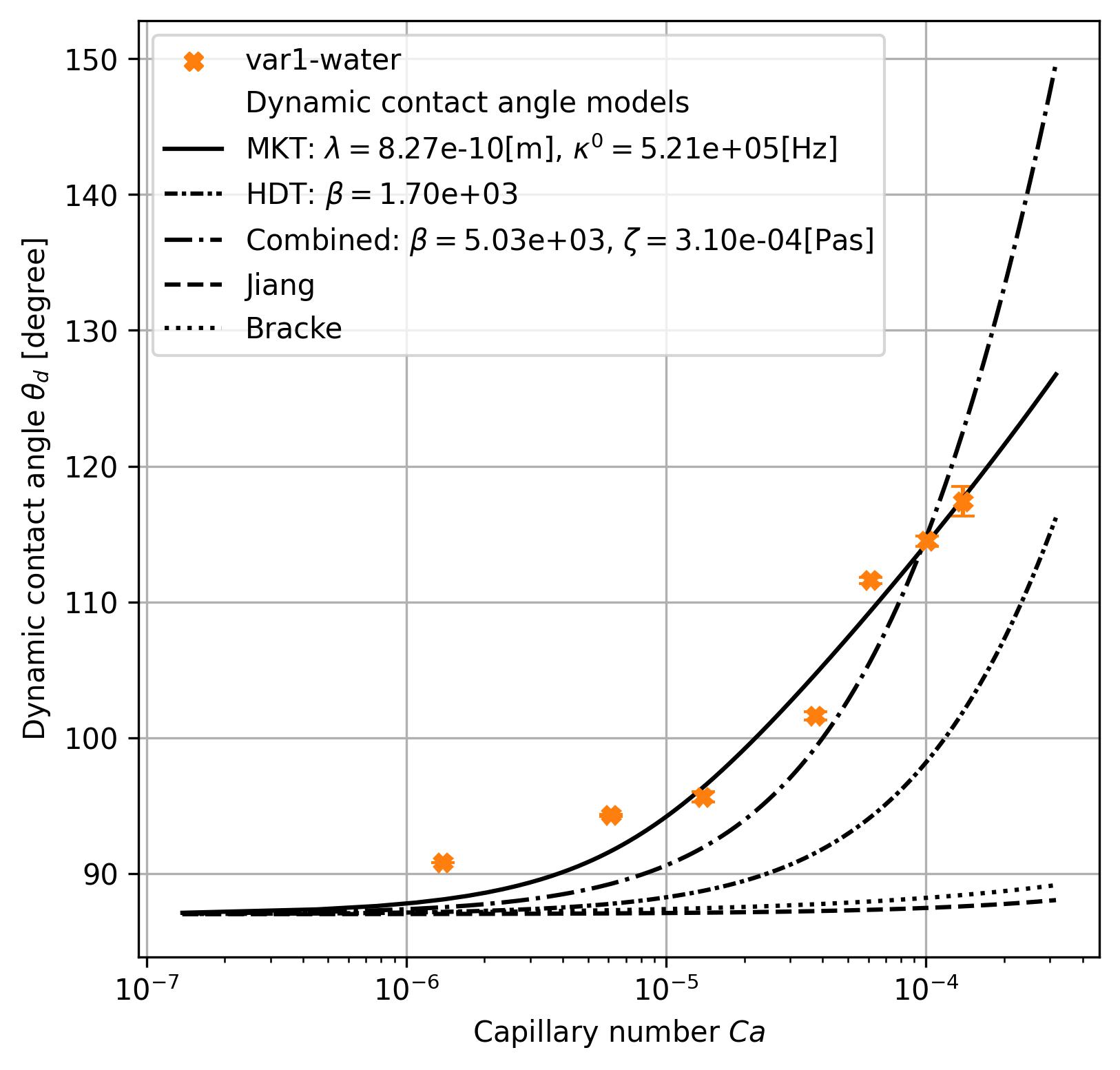}
        \caption{Variation 1}
        \label{fig:5.3.2_var1}
    \end{subfigure}
    \hfill
    \begin{subfigure}[t]{.44\linewidth}
        \centering
        \includegraphics[width=\linewidth]{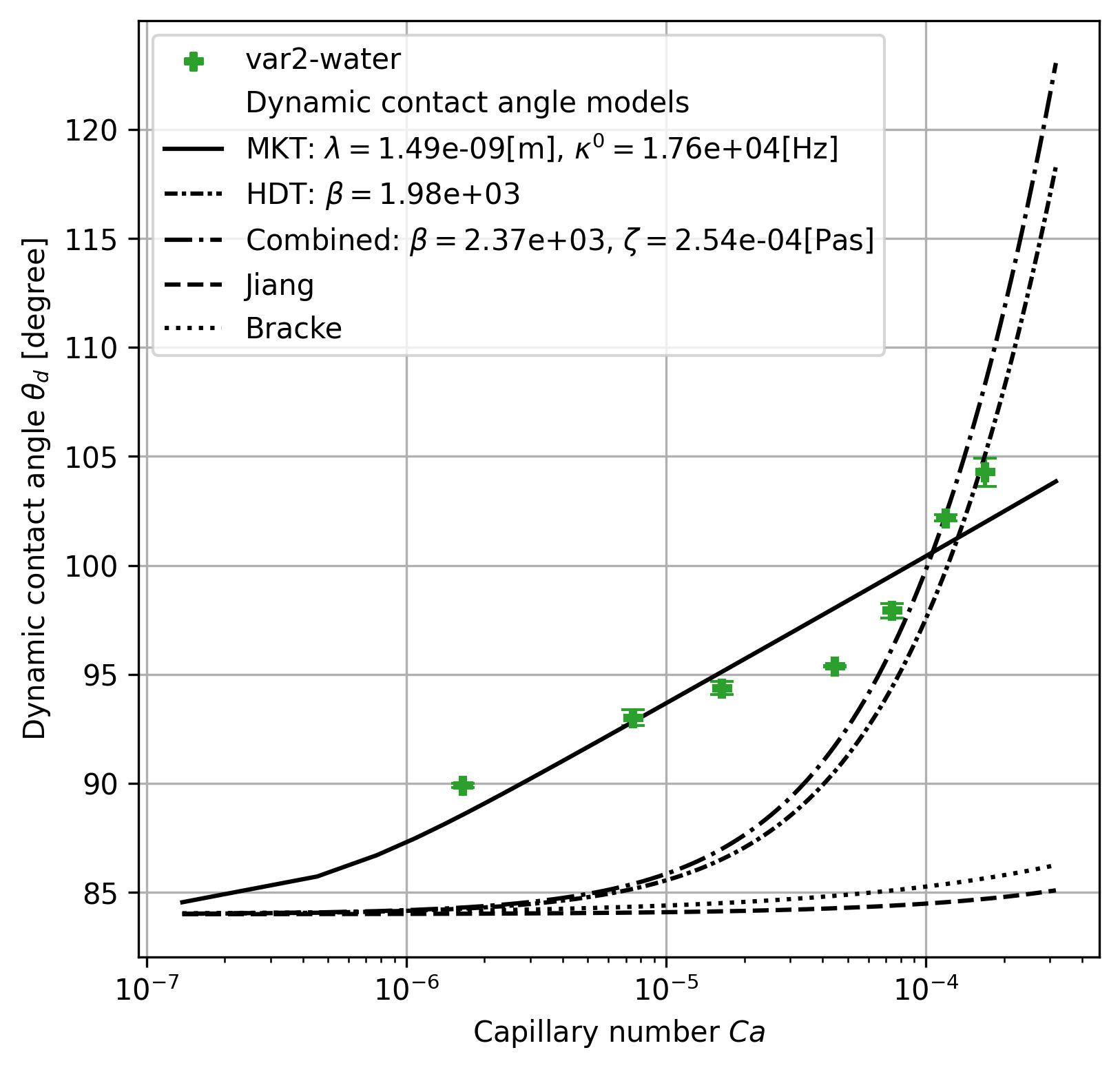}
        \caption{Variation 2}
        \label{fig:5.3.3_var2}
    \end{subfigure}
    \hfill
    \begin{subfigure}[t]{.44\linewidth}
        \centering
        \includegraphics[width=\linewidth]{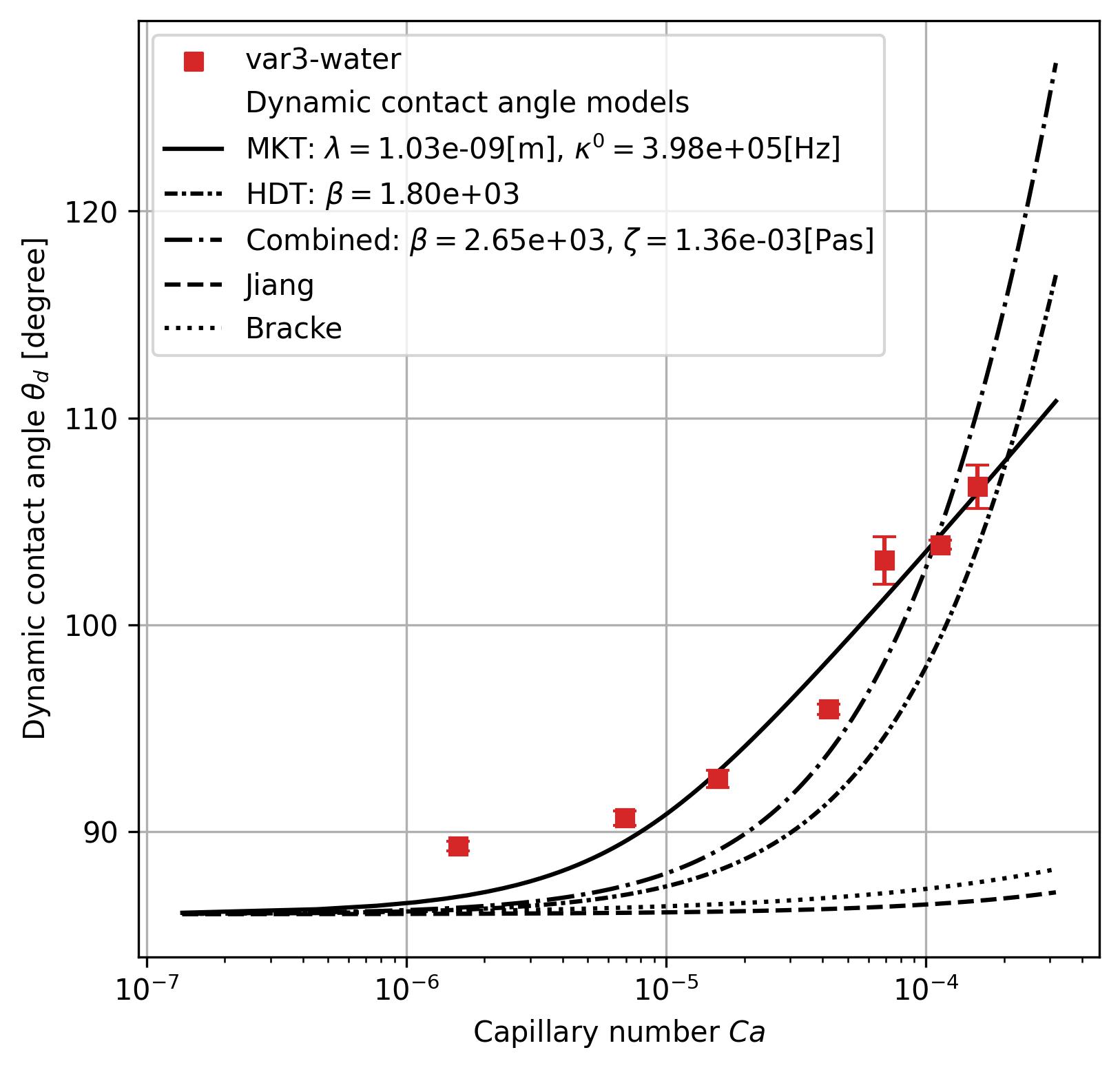}
        \caption{Variation 3}
        \label{fig:5.3.4_var3}
    \end{subfigure}
\caption{Comparison of five dynamic contact angle models with flow through four microchannels with free parameters estimated via curve-fitting.}
\label{fig:5.3_modelsComparison}
\end{figure*}

As stated in \cref{sec:02_dynamicContactAngle}, five dynamic CA models given in \cref{tab:2.1_dynCA_models} are compared with the experimental data from the forced wetting through various microchannels and the results are presented in \cref{fig:5.3_modelsComparison}. The free parameters for MKT, HDT and the combined model obtained from curve-fitting are displayed on each plot, respectively. The curve-fitting is realized by applying the \textit{scipy.optimize.curve\_fit} function developed by \cite{2020SciPy-NMeth} on the experimental data. The MKT stands out with the best performance among all models. The HDT and combined models underestimate the CA mainly for small $Ca$ numbers with even physical unreasonable parameters, while the empirical models tend to drastically underestimate the advancing dynamic CA in all here present regimes. To quantify the model performance, a model error indicator $\sigma$ is introduced as
\begin{equation}
    \sigma = \sqrt{\frac{\sum_{i=0}^n (\theta_{\text{exp},i} - \theta_{\text{model},i})^2}{n}}
\end{equation}
with $\theta_{exp}$ representing the dynamic CA from experimental measurements, $\theta_{model}$ the dynamic CA calculated from different models and $n$ the number of flow rates. The results are summarized in \cref{tab:5.1_modelError}, showing the low error of MKT. Consequently, the MKT model is chosen for further analysis out of the five models for the reason of outperforming.

\begin{table*}[!htb]
\centering
\caption{The standard deviation of five dynamic contact angle models in comparison with experimental results.}
\label{tab:5.1_modelError}
\begin{tabular}{c c c c c c}
\toprule[1.5pt]
Microchannel - Fluid & $\sigma_{\text{MKT}}$ & $\sigma_{\text{HDT}}$ & $\sigma_{\text{Combined}}$ & $\sigma_{\text{Jiang}}$ & $\sigma_{\text{Bracke}}$ \\
 & (deg) & (deg) & (deg) & (deg) & (deg) \\
\midrule[0.5pt]
Straight - 50\% Glycerin-water & 2.24 & 7.20 & 4.19 & 13.89 & 12.44 \\
Variation 1 - Water & 2.95 & 12.60 & 4.22 & 20.05 & 19.48 \\
Variation 2 - Water & 1.72 & 5.33 & 5.03 & 13.2 & 12.59 \\
Variation 3 - Water & 1.76 & 4.87 & 3.32 & 12.83 & 12.25 \\
\bottomrule[1.5pt]
\end{tabular}
\end{table*}

The static CA $\theta_0$ and the values for the fitting parameters with standard deviations are summarized in \cref{tab:5.2_mkt_geometry}, whose orders of magnitude are physically reasonable (see \cref{sec:02_dynamicContactAngle}). Despite the curved parts of the microchannel variation 1 and variation 2, their curvature ratio $R_o/R_i = 1.09$ is small and could be utilized directly for the comparison with straight channel. Nevertheless, the curved part of variation 3 with $R_o/R_i = 5.29$ could not be ignored, which results in the extraction of the experimental data from the straight part (\cref{fig:5.4.1_straight}) of variation 3 for the further analysis.

 For both working fluids, the measured dynamic CA matches well with the MKT theory, meaning the advancing dynamic CA increases with larger capillary number. That confirms the MKT's suitability for describing the CL dynamics of forced wetting through straight microchannels, not only limited by the conventional experimental methods (for example, Wilhelmy plate investigated by \cite{10.1063/5.0102028}). What stands out in \cref{tab:5.2_mkt_geometry} is the increase of $\kappa^0$ with more viscous fluid. That indicates that the molecules of more viscous fluid tend to attach and detach faster than less viscous fluid, which appears counter-intuitive. According to the definition of the equilibrium frequency in \cref{sec:02_dynamicContactAngle}, the different $\kappa^0$ could be related to, for example, the fluid viscosity and the friction coefficient affected by the local roughness (\cite{doi:10.1021/la4017917}; \cite{doi:10.1021/la202836q}).

Extending the forced flow investigation through the straight channel to three more different microchannels, as introduced in \cref{fig:3.1_samples}, the scope of the MKT's appropriateness for forced wetting could be broadened. It can be seen that the measured dynamic CAs depend on the channel geometry. For each geometry, a separate curve-fitting is performed and the obtained parameters are given in \cref{tab:5.2_mkt_geometry}. Interestingly, even with the same working fluid, the equilibrium frequency $\kappa^0$ varies from \SI{11.94}{kHz} to \SI{756.56}{kHz}, which might be attributed to the non-negligible local effects caused by the locally varied heat transfer to the bulk material (PMMA) during milling process. This might result in an non-uniform surface roughness and in consequence to a different CA.

\begin{table*}[!htb]
\centering
\caption{The static contact angle and the MKT parameters estimated from curve-fitting analysis for 50 wt\% Glycerin/water mixture through straight channel and water through three curved microchannels.}
\label{tab:5.2_mkt_geometry}
\begin{tabular}{c c c c}
\toprule[1.5pt]
Microchannel - Fluid & $\theta_0$ (deg) & $\lambda$ (\SI{}{nm}) & $\kappa^0$ (\SI{}{kHz}) \\
\midrule[0.5pt]
Straight - 50\% Glycerin-water & 82 & 1.17 $\pm$ 0.05 & 220.54 $\pm$ 76.41 \\
Variation 1 - Water & 87 & 0.83 $\pm$ 0.04 & 521.13 $\pm$ 137.69 \\
Variation 2 - Water & 84 & 1.48 $\pm$ 0.07 & 17.64 $\pm$ 8.42 \\
Variation 3 - Water & 86 & 1.03 $\pm$ 0.05 & 397.82 $\pm$ 128.46 \\
\bottomrule[1.5pt]
\end{tabular}
\end{table*}

\subsection{Influence of curved microchannel on the dynamic contact angle}
\label{subsec:5.3_curvetureRatio}

Now that the MKT's suitability is confirmed by two working fluids and four different microchannels, the performance of MKT with forced flow through microchannel with large curvature ratio is further studied. 

\begin{figure}[!htb]
\centering
    \begin{subfigure}[t]{0.8\linewidth}
        \centering
        \includegraphics[width=\linewidth]{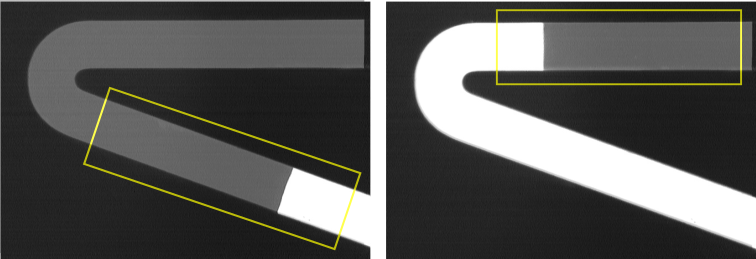}
        \caption{Straight part (yellow). Left: $t=\SI{0.983}{s}$. Right: $t=\SI{10.716}{s}$.}
        \label{fig:5.4.1_straight}
    \end{subfigure}
    \hfill
    \begin{subfigure}[t]{0.8\linewidth}
        \centering
        \includegraphics[width=\linewidth]{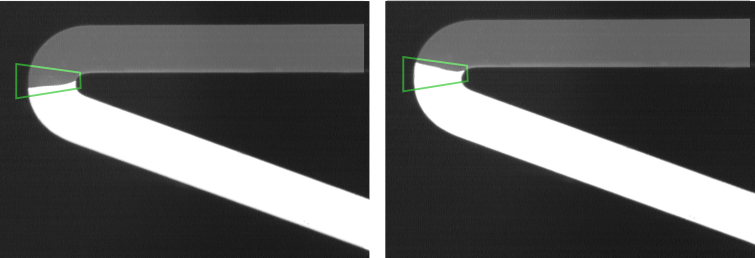}
        \caption{Position 1 (green). Left: $t=\SI{7.516}{s}$. Right: $t=\SI{7.783}{s}$.}
        \label{fig:5.4.2_position1}
    \end{subfigure}
    \hfill
    \begin{subfigure}[t]{0.8\linewidth}
        \centering
        \includegraphics[width=\linewidth]{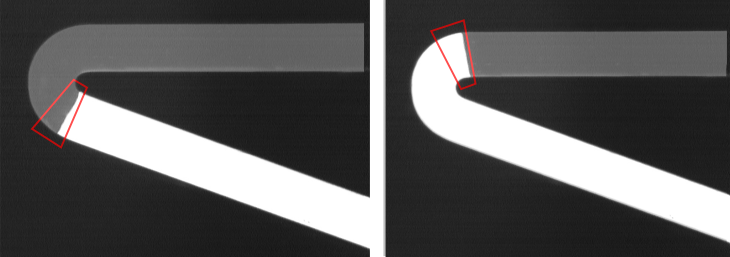}
        \caption{Position 2 (red). Left: $t=\SI{6.716}{s}$. Right: $t=\SI{8.850}{s}$.}
        \label{fig:5.4.3_position2}
    \end{subfigure}
\caption{The microchannel variation 3 is classified as three parts ($\dot{V} = \SI{0.00011}{mL/s}$): (a) straight part (yellow), (b) position 1 (green) and (c) position 2 (red).}
\label{fig:5.4_var3_parts}
\end{figure}

To study how the curved part will affect the dynamic wetting behavior, the microchannel variation 3 is divided into two parts: straight part (\cref{fig:5.4.1_straight}) and curved part, while the curved part is further classified as position 1 (\cref{fig:5.4.2_position1}) and position 2 (\cref{fig:5.4.3_position2}) for the reason of exhibiting different wetting behaviors. The position 1 refers to the middle of the curved part, while the position 2 covers the begin and the end. Owing to the large curvature ratio $R_o/R_i = 5.29$, the dynamic CAs of curved part are measured at inside and outside, respectively. Having the information of inside radius $R_i$, outside radius $R_o$ (\cref{tab:3.1_channelGeometry}) and the flow velocity allows the calculations of local CL velocities at the inside and the outside of curved parts.

\begin{figure}[!htb]
    \centering
    \includegraphics[width=\linewidth]{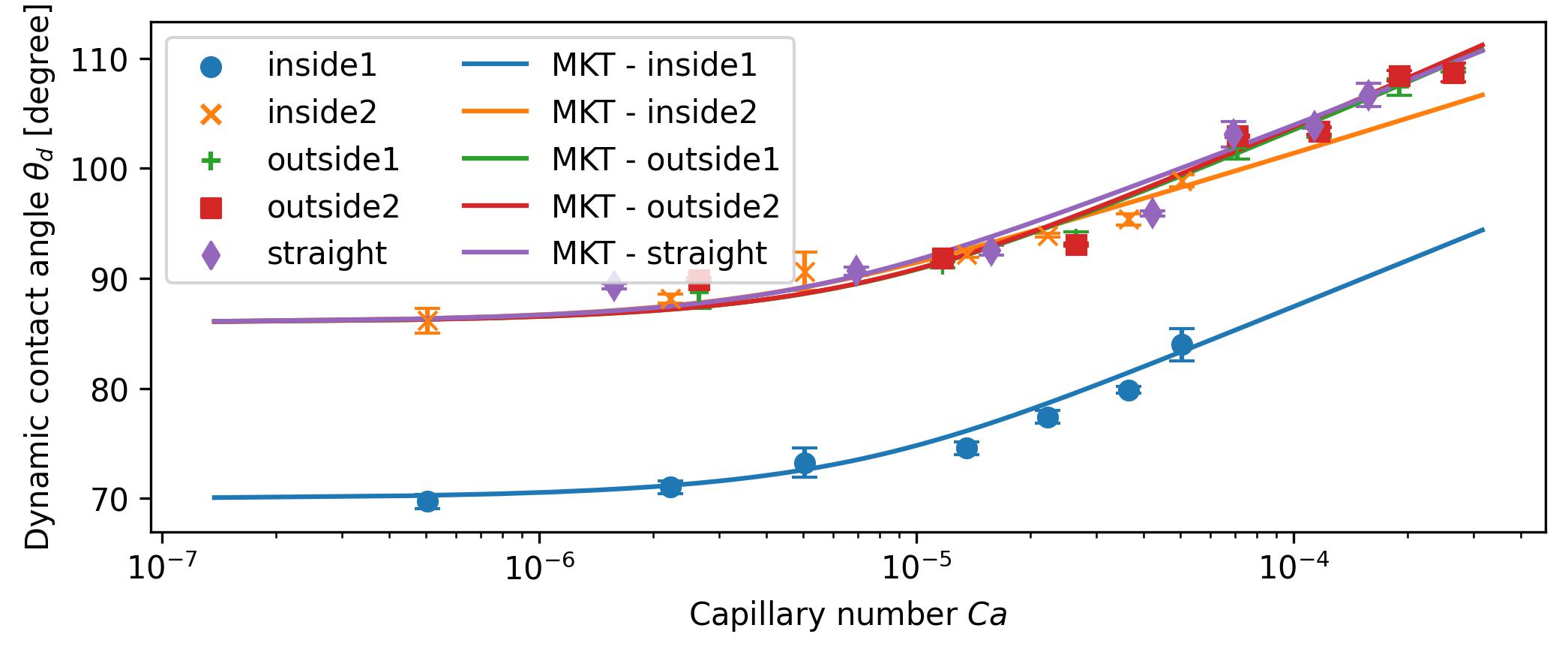}
    \caption{Dynamic contact angle $\theta_d$ versus capillary number $\mathit{Ca}$ with water through three parts of variation 3: straight part, position\,1 and position\,2.}
    \label{fig:5.5_mkt_var3}
\end{figure}

Plotting the dynamic CA $\theta_d$ with $\mathit{Ca}$ calculated using local CL velocities, the comparison between experimental results and fit to the MKT model can be seen in \cref{fig:5.5_mkt_var3}. The free parameters $\kappa^0$ and $\lambda$ estimated from curve-fitting with standard deviations are listed in \cref{tab:5.3_mkt_var3}. It should be noted that the identical static CA $\theta_0 = 86$° is employed for all parts of variation 3, except for the inside of position 1, where $\theta_0 = 70$° is applied. That is attributed to the observed significantly small dynamic CAs, resulting from the formation of a distinct tip between the interface and the wall as presented in \cref{fig:5.4.2_position1}. This strong bending of interface along the inside CL is mostly likely due to the considerable difference of the local CL velocities between the inner and the outer radius of the microchannel, causing the pinning of the inner radius CL. The different shear rates on curved region might also contribute to this phenomenon. In addition, it is worth stressing that the rectangular geometry of a channel is known to cause the ``finger" effect of the CL along the sharp edges of channel (\cite{DONG1995278}; \cite{WONG1992317}), which might be enhanced by the pronounced curvature of variation 3.

As a consequence of the smaller dynamic CA at the inside of position 1 than the measured static CA $\theta_0=86$°, a derived static CA $\theta_0 = 70$° from the previous results is utilized for the inside of position 1. As shown in \cref{fig:5.5_mkt_var3}, the dynamic wetting behavior of water through variation 3 could be well interpreted using MKT, with the obtained parameters (\cref{tab:5.3_mkt_var3}) not only physically meaningful, but also lay in a similar range. Again, the suitability of the MKT approach to describe the CL dynamics of forced wetting through microchannels is confirmed, even in scenarios involving geometrically complex microchannels. The dynamic CA depends not only on the channel geometry, but also significantly on the curvature ratio.

\begin{table}[h]
\centering
\caption{The static contact angle and the MKT parameters estimated from curve-fitting analysis for forced wetting through three parts of variation 3 with water.}
\label{tab:5.3_mkt_var3}
\begin{tabular}{c c c c}
\toprule[1.5pt]
Channel part & $\theta_0$ & $\lambda$ & $\kappa^0$ \\
 & (deg) & (\SI{}{nm}) & (\SI{}{kHz}) \\
\midrule[0.5pt]
Straight & 86 & 1.03 $\pm$ 0.05 & 385.61 $\pm$ 124.52 \\
Inside1 & 70 & 1.03 $\pm$ 0.11 & 424.12 $\pm$ 153.90 \\
Outside1 & 86 & 1.02 $\pm$ 0.03 & 457.26 $\pm$ 90.52 \\
Inside2 & 86 & 1.34 $\pm$ 0.09 & 107.37 $\pm$ 43.47 \\
Outside2 & 86 & 1.03 $\pm$ 0.04 & 406.92 $\pm$ 111.90 \\
\bottomrule[1.5pt]
\end{tabular}
\end{table}
%

\section{Conclusions and Outlook}
\label{sec:06_conclusions}

To gain a better understanding of the sealing issue in industrial products, where the leakage pots are prone to posses rough surface and complex geometries, an appropriate dynamic contact angle (CA) model is crucial. In present work, the dynamic wetting behaviors of forced flow with small capillary number $\mathit{Ca}$ ($10^{-6}$ - $10^{-3}$) through geometrically complex microchannels are investigated experimentally, to cover the most common leaking circumstances in application.  In addition to the industrial most relevant fluid - water, 50 wt\% Glycerin/water mixture is employed to examine the suitability of MKT approach. The forced wetting through three curved microchannels with different geometry and curvature ratio $R_o/R_i$ is conducted and the findings confirm again the feasibility of applying the MKT on the CL dynamics involved in industrial circumstances, meaning not only the flow through microchannels with rough surface, but also with geometrical complexity. The channel geometry as well as the significantly different curvatures of curved microchannels have influence on the CL movement. However, the deviation of the free parameter $\kappa^0$ between different experiments is still not clear. The other phenomenon need to be further explored is the extremely small dynamic CAs appearing on the inside of the position 1 of variation 3. More studies are suggested to include curved channels with ratios between 1.09 and 5.23, so that the pinning effect on this position could be further accompanied by numerical simulations to investigate the flow fields near the inner CL.

Moreover, the proposed automated image analysis for dynamic CA measurement provides a straightforward and reliable means to systematically analyze the interface movement within images acquired from non-transparent samples, meaning the interface could not be extracted easily. It not only offers simplicity in its application but significantly reduces the time invested in manual measurements. 

The presented experimental results will serve as a basis for the further 3D transient numerical investigations of dynamic wetting behaviors to compliment the 2D experimental results and cover a more broad spectrum of industrial media tightness topic.

\section{Acknowledgments}

The authors would like to thank Prof. Dr. Joël De Coninck of Université libre de Bruxelles for the helpful and valuable discussion on the experimental results. 
The last two authors acknowledge the funding by the German Research Foundation (DFG): July 1 2020 - 30 June 2024
funded by the German Research Foundation (DFG) - Project-ID 265191195 - SFB 1194.

\appendix
\section{Appendix: Rectangular channel geometry and corner fluid rivulet}
\label{appdendice1}

\begin{figure}[!htb]
    \centering
    \includegraphics[width=0.5\textwidth]{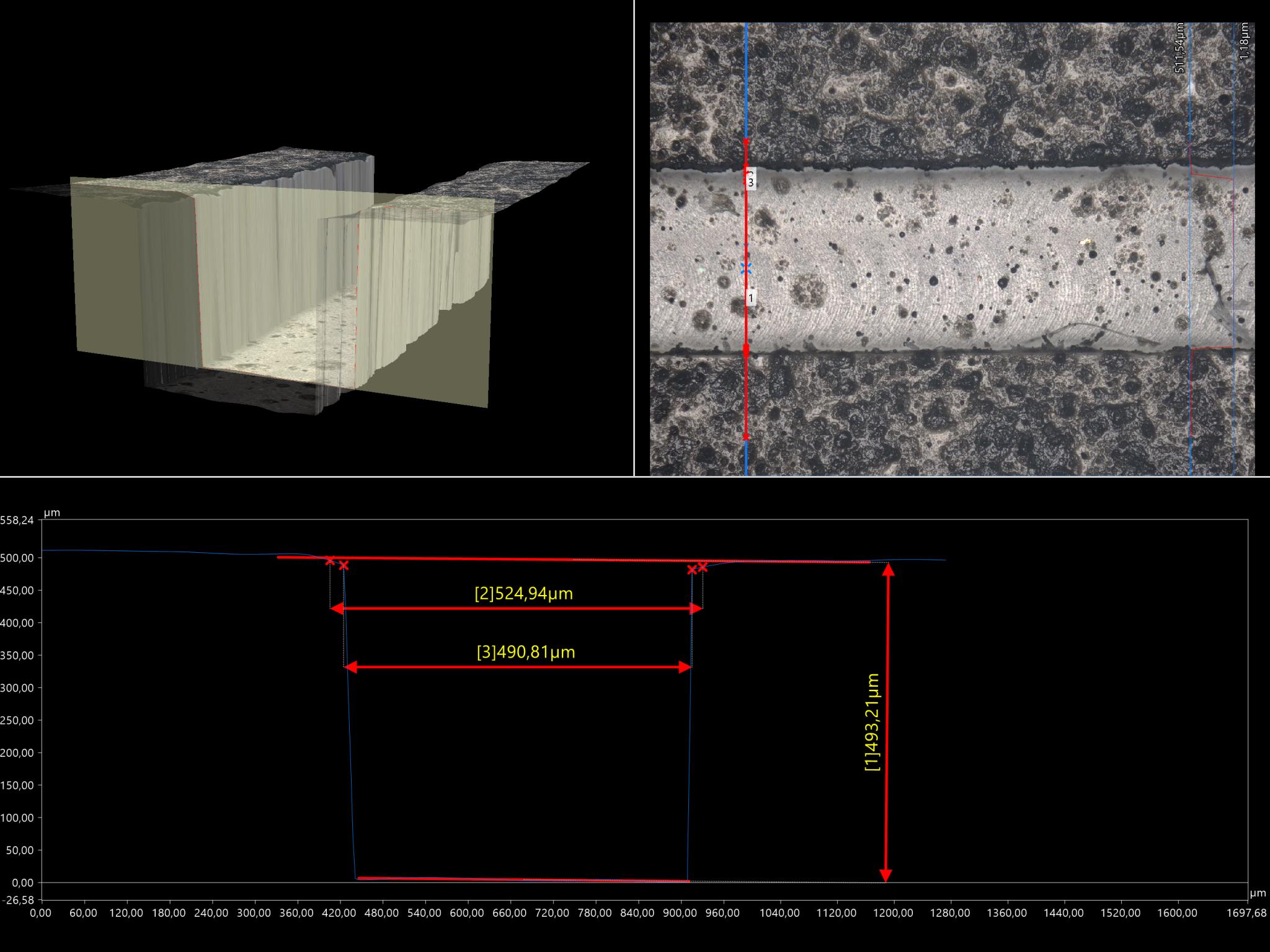}
    \caption{Microscopy of the straight channel cross-section. Top left: depth profile of the chosen section of straight channel. Top right: top view of the chosen-section and the chose position for measurements (blue line). Bottom: depth profile of the chosen position and measurements (red line).}
    \label{fig:app_microscopy}
\end{figure}

In \cref{fig:app_microscopy} the microscopy of the cross-section of the straight channel (\cref{fig:3.1.1_straight}) is displayed, showing the exact channel dimension of the chosen position and the sharp edge of the channel by micromilling.

\begin{figure}[!htb]
\centering
\def\svgwidth{0.5\textwidth}
\begingroup%
  \makeatletter%
  \providecommand\color[2][]{%
    \errmessage{(Inkscape) Color is used for the text in Inkscape, but the package 'color.sty' is not loaded}%
    \renewcommand\color[2][]{}%
  }%
  \providecommand\transparent[1]{%
    \errmessage{(Inkscape) Transparency is used (non-zero) for the text in Inkscape, but the package 'transparent.sty' is not loaded}%
    \renewcommand\transparent[1]{}%
  }%
  \providecommand\rotatebox[2]{#2}%
  \newcommand*\fsize{\dimexpr\f@size pt\relax}%
  \newcommand*\lineheight[1]{\fontsize{\fsize}{#1\fsize}\selectfont}%
  \ifx\svgwidth\undefined%
    \setlength{\unitlength}{329.57972882bp}%
    \ifx\svgscale\undefined%
      \relax%
    \else%
      \setlength{\unitlength}{\unitlength * \real{\svgscale}}%
    \fi%
  \else%
    \setlength{\unitlength}{\svgwidth}%
  \fi%
  \global\let\svgwidth\undefined%
  \global\let\svgscale\undefined%
  \makeatother%
  \begin{picture}(1,1.09616653)%
    \lineheight{1}%
    \setlength\tabcolsep{0pt}%
    \put(0,0){\includegraphics[width=\unitlength,page=1]{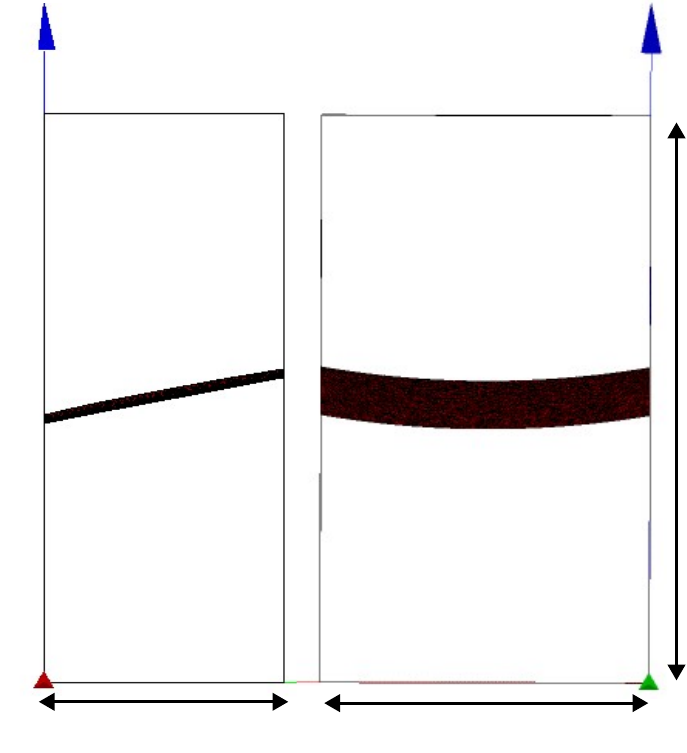}}%
    \put(0.04116463,0.01068298){\color[rgb]{0,0,0}\makebox(0,0)[lt]{\lineheight{1.25}\smash{\begin{tabular}[t]{l}Depth = 0.406mm\end{tabular}}}}%
    \put(0.5177583,0.01068298){\color[rgb]{0,0,0}\makebox(0,0)[lt]{\lineheight{1.25}\smash{\begin{tabular}[t]{l}Width = 0.554mm\end{tabular}}}}%
    \put(0.62117191,0.84181558){\color[rgb]{0,0,0}\makebox(0,0)[lt]{\lineheight{1.25}\smash{\begin{tabular}[t]{l}Length = 1mm\end{tabular}}}}%
    \put(0.11836898,0.40730771){\color[rgb]{0,0,0}\makebox(0,0)[lt]{\lineheight{1.25}\smash{\begin{tabular}[t]{l}TPU\\$\theta_0=104$°\end{tabular}}}}%
    \put(0.24112234,0.64011553){\color[rgb]{0,0,0}\makebox(0,0)[lt]{\lineheight{1.25}\smash{\begin{tabular}[t]{l}PMMA\\$\theta_0 = 80$°\end{tabular}}}}%
  \end{picture}%
\endgroup%

\caption{Simulation result in Surface Evolver with three PMMA walls ($\theta_0=80$°) and one TPU wall ($\theta_0 = 104$°). Below the interface is water while above the interface is air. Left: side view. Right: top view.}
\label{fig:app_se}
\end{figure}

It is well-known that the wetting in a rectangular channel leads to rivulets along the sharp corner (\cite{KUBOCHKIN2022101575}; \cite{PhysRevFluids.7.114002}), however, \cite{doi:10.1073/pnas.63.2.292} found that for the wetting cases with an equilibrium contact angle $\theta_0>45$°, there are no rivulets. Employing the 3D Surface Evolver \citep{doi:10.1080/10586458.1992.10504253} as an addition source for the microchannel with three PMMA walls ($\theta_0=80$°) and one TPU wall ($\theta_0 = 104$°), the resolved interface in channel with the exact 2D top-down view as in experiment is given in \cref{fig:app_se}. With no visible corner rivulet in \cref{fig:app_se}, it can be said that the results obtained with a 2D top-down view from experiments are not affected by rivulets.





\bibliography{sn-bibliography}


\begin{thebibliography}{56}
\ifx \bisbn   \undefined \def \bisbn  #1{ISBN #1}\fi
\ifx \binits  \undefined \def \binits#1{#1}\fi
\ifx \bauthor  \undefined \def \bauthor#1{#1}\fi
\ifx \batitle  \undefined \def \batitle#1{#1}\fi
\ifx \bjtitle  \undefined \def \bjtitle#1{#1}\fi
\ifx \bvolume  \undefined \def \bvolume#1{\textbf{#1}}\fi
\ifx \byear  \undefined \def \byear#1{#1}\fi
\ifx \bissue  \undefined \def \bissue#1{#1}\fi
\ifx \bfpage  \undefined \def \bfpage#1{#1}\fi
\ifx \blpage  \undefined \def \blpage #1{#1}\fi
\ifx \burl  \undefined \def \burl#1{\textsf{#1}}\fi
\ifx \doiurl  \undefined \def \doiurl#1{\url{https://doi.org/#1}}\fi
\ifx \betal  \undefined \def \betal{\textit{et al.}}\fi
\ifx \binstitute  \undefined \def \binstitute#1{#1}\fi
\ifx \binstitutionaled  \undefined \def \binstitutionaled#1{#1}\fi
\ifx \bctitle  \undefined \def \bctitle#1{#1}\fi
\ifx \beditor  \undefined \def \beditor#1{#1}\fi
\ifx \bpublisher  \undefined \def \bpublisher#1{#1}\fi
\ifx \bbtitle  \undefined \def \bbtitle#1{#1}\fi
\ifx \bedition  \undefined \def \bedition#1{#1}\fi
\ifx \bseriesno  \undefined \def \bseriesno#1{#1}\fi
\ifx \blocation  \undefined \def \blocation#1{#1}\fi
\ifx \bsertitle  \undefined \def \bsertitle#1{#1}\fi
\ifx \bsnm \undefined \def \bsnm#1{#1}\fi
\ifx \bsuffix \undefined \def \bsuffix#1{#1}\fi
\ifx \bparticle \undefined \def \bparticle#1{#1}\fi
\ifx \barticle \undefined \def \barticle#1{#1}\fi
\bibcommenthead
\ifx \bconfdate \undefined \def \bconfdate #1{#1}\fi
\ifx \botherref \undefined \def \botherref #1{#1}\fi
\ifx \url \undefined \def \url#1{\textsf{#1}}\fi
\ifx \bchapter \undefined \def \bchapter#1{#1}\fi
\ifx \bbook \undefined \def \bbook#1{#1}\fi
\ifx \bcomment \undefined \def \bcomment#1{#1}\fi
\ifx \oauthor \undefined \def \oauthor#1{#1}\fi
\ifx \citeauthoryear \undefined \def \citeauthoryear#1{#1}\fi
\ifx \endbibitem  \undefined \def \endbibitem {}\fi
\ifx \bconflocation  \undefined \def \bconflocation#1{#1}\fi
\ifx \arxivurl  \undefined \def \arxivurl#1{\textsf{#1}}\fi
\csname PreBibitemsHook\endcsname

\bibitem[\protect\citeauthoryear{Bracke et~al.}{1989}]{10.1007/BFb0116200}
\begin{bchapter}
\bauthor{\bsnm{Bracke}, \binits{M.}},
\bauthor{\bsnm{De~Voeght}, \binits{F.}},
\bauthor{\bsnm{Joos}, \binits{P.}}:
\bctitle{The kinetics of wetting: the dynamic contact angle}.
In: \beditor{\bsnm{Bothorel}, \binits{P.}},
\beditor{\bsnm{Dufourc}, \binits{E.J.}} (eds.)
\bbtitle{Trends in Colloid and Interface Science III},
pp. \bfpage{142}--\blpage{149}.
\bpublisher{Steinkopff},
\blocation{Darmstadt}
(\byear{1989})
\end{bchapter}
\endbibitem

\bibitem[\protect\citeauthoryear{Bonn et~al.}{2009}]{Bonn2009739}
\begin{barticle}
\bauthor{\bsnm{Bonn}, \binits{D.}},
\bauthor{\bsnm{Eggers}, \binits{J.}},
\bauthor{\bsnm{Indekeu}, \binits{J.}},
\bauthor{\bsnm{Meunier}, \binits{J.}}:
\batitle{Wetting and spreading}.
\bjtitle{Reviews of Modern Physics}
\bvolume{81}(\bissue{2}),
\bfpage{739}--\blpage{805}
(\byear{2009})
\doiurl{10.1103/RevModPhys.81.739} .
\bcomment{Cited by: 2190; All Open Access, Green Open Access}
\end{barticle}
\endbibitem

\bibitem[\protect\citeauthoryear{Blake and Haynes}{1969}]{BLAKE1969421}
\begin{barticle}
\bauthor{\bsnm{Blake}, \binits{T.D.}},
\bauthor{\bsnm{Haynes}, \binits{J.M.}}:
\batitle{Kinetics of liquidliquid displacement}.
\bjtitle{Journal of Colloid and Interface Science}
\bvolume{30}(\bissue{3}),
\bfpage{421}--\blpage{423}
(\byear{1969})
\doiurl{10.1016/0021-9797(69)90411-1}
\end{barticle}
\endbibitem

\bibitem[\protect\citeauthoryear{Blake}{1969}]{BlakeDiss}
\begin{botherref}
\oauthor{\bsnm{Blake}, \binits{T.D.}}:
The contact angle and two-phase flow.
PhD thesis,
The University of BristolSchool of Chemistry
(1969).
\url{https://research-information.bris.ac.uk/en/studentTheses/the-contact-angle-and-two-phase-flow}
\end{botherref}
\endbibitem

\bibitem[\protect\citeauthoryear{Blake}{1993}]{Blake1993}
\begin{botherref}
\oauthor{\bsnm{Blake}, \binits{T.D.}}:
Dynamic contact angle and wetting kinetics.
Wettability
(1993)
\end{botherref}
\endbibitem

\bibitem[\protect\citeauthoryear{Blake}{2006}]{BLAKE20061}
\begin{barticle}
\bauthor{\bsnm{Blake}, \binits{T.D.}}:
\batitle{The physics of moving wetting lines}.
\bjtitle{Journal of Colloid and Interface Science}
\bvolume{299}(\bissue{1}),
\bfpage{1}--\blpage{13}
(\byear{2006})
\doiurl{10.1016/j.jcis.2006.03.051}
\end{barticle}
\endbibitem

\bibitem[\protect\citeauthoryear{Brakke}{1992}]{doi:10.1080/10586458.1992.10504253}
\begin{barticle}
\bauthor{\bsnm{Brakke}, \binits{K.A.}}:
\batitle{The surface evolver}.
\bjtitle{Experimental Mathematics}
\bvolume{1}(\bissue{2}),
\bfpage{141}--\blpage{165}
(\byear{1992})
\doiurl{10.1080/10586458.1992.10504253}
{\href{https://arxiv.org/abs/https://doi.org/10.1080/10586458.1992.10504253}{{https://doi.org/10.1080/10586458.1992.10504253}}}
\end{barticle}
\endbibitem

\bibitem[\protect\citeauthoryear{Brochard-Wyart and {de
  Gennes}}{1992}]{BROCHARDWYART19921}
\begin{barticle}
\bauthor{\bsnm{Brochard-Wyart}, \binits{F.}},
\bauthor{\bsnm{{de Gennes}}, \binits{P.G.}}:
\batitle{Dynamics of partial wetting}.
\bjtitle{Advances in Colloid and Interface Science}
\bvolume{39},
\bfpage{1}--\blpage{11}
(\byear{1992})
\doiurl{10.1016/0001-8686(92)80052-Y}
\end{barticle}
\endbibitem

\bibitem[\protect\citeauthoryear{Cahn}{1977}]{Cahn}
\begin{barticle}
\bauthor{\bsnm{Cahn}, \binits{J.W.}}:
\batitle{{Critical point wetting}}.
\bjtitle{The Journal of Chemical Physics}
\bvolume{66}(\bissue{8}),
\bfpage{3667}--\blpage{3672}
(\byear{1977})
\doiurl{10.1063/1.434402}
{\href{https://arxiv.org/abs/https://pubs.aip.org/aip/jcp/article-pdf/66/8/3667/18906009/3667\_1\_online.pdf}{{https://pubs.aip.org/aip/jcp/article-pdf/66/8/3667/18906009/3667\_1\_online.pdf}}}
\end{barticle}
\endbibitem

\bibitem[\protect\citeauthoryear{Chen
  et~al.}{2017}]{doi:10.1021/acs.langmuir.7b01223}
\begin{barticle}
\bauthor{\bsnm{Chen}, \binits{X.}},
\bauthor{\bsnm{Chen}, \binits{J.}},
\bauthor{\bsnm{Ouyang}, \binits{X.}},
\bauthor{\bsnm{Song}, \binits{Y.}},
\bauthor{\bsnm{Xu}, \binits{R.}},
\bauthor{\bsnm{Jiang}, \binits{P.}}:
\batitle{Water droplet spreading and wicking on nanostructured surfaces}.
\bjtitle{Langmuir}
\bvolume{33}(\bissue{27}),
\bfpage{6701}--\blpage{6707}
(\byear{2017})
\doiurl{10.1021/acs.langmuir.7b01223}
{\href{https://arxiv.org/abs/https://doi.org/10.1021/acs.langmuir.7b01223}{{https://doi.org/10.1021/acs.langmuir.7b01223}}}.
\bcomment{PMID: 28609626}
\end{barticle}
\endbibitem

\bibitem[\protect\citeauthoryear{Concus and
  Finn}{1969}]{doi:10.1073/pnas.63.2.292}
\begin{barticle}
\bauthor{\bsnm{Concus}, \binits{P.}},
\bauthor{\bsnm{Finn}, \binits{R.}}:
\batitle{On the behavior of a capillary surface in a wedge<sup>*</sup>}.
\bjtitle{Proceedings of the National Academy of Sciences}
\bvolume{63}(\bissue{2}),
\bfpage{292}--\blpage{299}
(\byear{1969})
\doiurl{10.1073/pnas.63.2.292}
{\href{https://arxiv.org/abs/https://www.pnas.org/doi/pdf/10.1073/pnas.63.2.292}{{https://www.pnas.org/doi/pdf/10.1073/pnas.63.2.292}}}
\end{barticle}
\endbibitem

\bibitem[\protect\citeauthoryear{Cahn and Hilliard}{1958}]{Cahn_Hilliard}
\begin{barticle}
\bauthor{\bsnm{Cahn}, \binits{J.W.}},
\bauthor{\bsnm{Hilliard}, \binits{J.E.}}:
\batitle{{Free Energy of a Nonuniform System. I. Interfacial Free Energy}}.
\bjtitle{The Journal of Chemical Physics}
\bvolume{28}(\bissue{2}),
\bfpage{258}--\blpage{267}
(\byear{1958})
\doiurl{10.1063/1.1744102}
{\href{https://arxiv.org/abs/https://pubs.aip.org/aip/jcp/article-pdf/28/2/258/18813541/258\_1\_online.pdf}{{https://pubs.aip.org/aip/jcp/article-pdf/28/2/258/18813541/258\_1\_online.pdf}}}
\end{barticle}
\endbibitem

\bibitem[\protect\citeauthoryear{Cox}{1986}]{Cox1986169}
\begin{barticle}
\bauthor{\bsnm{Cox}, \binits{R.G.}}:
\batitle{The dynamics of the spreading of liquids on a solid surface. part 1.
  viscous flow}.
\bjtitle{Journal of Fluid Mechanics}
\bvolume{168},
\bfpage{169}--\blpage{194}
(\byear{1986})
\doiurl{10.1017/S0022112086000332} .
\bcomment{Cited by: 1210}
\end{barticle}
\endbibitem

\bibitem[\protect\citeauthoryear{Clanet and Quéré}{2002}]{clanet_quere_2002}
\begin{barticle}
\bauthor{\bsnm{Clanet}, \binits{C.}},
\bauthor{\bsnm{Quéré}, \binits{D.}}:
\batitle{Onset of menisci}.
\bjtitle{Journal of Fluid Mechanics}
\bvolume{460},
\bfpage{131}--\blpage{149}
(\byear{2002})
\doiurl{10.1017/S002211200200808X}
\end{barticle}
\endbibitem

\bibitem[\protect\citeauthoryear{Duvivier et~al.}{2013}]{doi:10.1021/la4017917}
\begin{barticle}
\bauthor{\bsnm{Duvivier}, \binits{D.}},
\bauthor{\bsnm{Blake}, \binits{T.D.}},
\bauthor{\bsnm{De~Coninck}, \binits{J.}}:
\batitle{Toward a predictive theory of wetting dynamics}.
\bjtitle{Langmuir}
\bvolume{29}(\bissue{32}),
\bfpage{10132}--\blpage{10140}
(\byear{2013})
\doiurl{10.1021/la4017917}
{\href{https://arxiv.org/abs/https://doi.org/10.1021/la4017917}{{https://doi.org/10.1021/la4017917}}}.
\bcomment{PMID: 23844877}
\end{barticle}
\endbibitem

\bibitem[\protect\citeauthoryear{Dong and Chatzis}{1995}]{DONG1995278}
\begin{barticle}
\bauthor{\bsnm{Dong}, \binits{M.}},
\bauthor{\bsnm{Chatzis}, \binits{I.}}:
\batitle{The imbibition and flow of a wetting liquid along the corners of a
  square capillary tube}.
\bjtitle{Journal of Colloid and Interface Science}
\bvolume{172}(\bissue{2}),
\bfpage{278}--\blpage{288}
(\byear{1995})
\doiurl{10.1006/jcis.1995.1253}
\end{barticle}
\endbibitem

\bibitem[\protect\citeauthoryear{de~Gennes}{1985}]{RevModPhys.57.827}
\begin{barticle}
\bauthor{\bsnm{Gennes}, \binits{P.G.}}:
\batitle{Wetting: statics and dynamics}.
\bjtitle{Rev. Mod. Phys.}
\bvolume{57},
\bfpage{827}--\blpage{863}
(\byear{1985})
\doiurl{10.1103/RevModPhys.57.827}
\end{barticle}
\endbibitem

\bibitem[\protect\citeauthoryear{de~Ruiter
  et~al.}{2017}]{PhysRevFluids.2.043602}
\begin{barticle}
\bauthor{\bsnm{Ruiter}, \binits{R.}},
\bauthor{\bsnm{Colinet}, \binits{P.}},
\bauthor{\bsnm{Brunet}, \binits{P.}},
\bauthor{\bsnm{Snoeijer}, \binits{J.H.}},
\bauthor{\bsnm{Gelderblom}, \binits{H.}}:
\batitle{Contact line arrest in solidifying spreading drops}.
\bjtitle{Phys. Rev. Fluids}
\bvolume{2},
\bfpage{043602}
(\byear{2017})
\doiurl{10.1103/PhysRevFluids.2.043602}
\end{barticle}
\endbibitem

\bibitem[\protect\citeauthoryear{Duvivier et~al.}{2011}]{doi:10.1021/la202836q}
\begin{barticle}
\bauthor{\bsnm{Duvivier}, \binits{D.}},
\bauthor{\bsnm{Seveno}, \binits{D.}},
\bauthor{\bsnm{Rioboo}, \binits{R.}},
\bauthor{\bsnm{Blake}, \binits{T.D.}},
\bauthor{\bsnm{De~Coninck}, \binits{J.}}:
\batitle{Experimental evidence of the role of viscosity in the molecular
  kinetic theory of dynamic wetting}.
\bjtitle{Langmuir}
\bvolume{27}(\bissue{21}),
\bfpage{13015}--\blpage{13021}
(\byear{2011})
\doiurl{10.1021/la202836q}
{\href{https://arxiv.org/abs/https://doi.org/10.1021/la202836q}{{https://doi.org/10.1021/la202836q}}}.
\bcomment{PMID: 21919445}
\end{barticle}
\endbibitem

\bibitem[\protect\citeauthoryear{Eddi et~al.}{2013}]{Eddi2013}
\begin{botherref}
\oauthor{\bsnm{Eddi}, \binits{A.}},
\oauthor{\bsnm{Winkels}, \binits{K.G.}},
\oauthor{\bsnm{Snoeijer}, \binits{J.H.}}:
Short time dynamics of viscous drop spreading.
Physics of Fluids
\textbf{25}(1)
(2013)
\doiurl{10.1063/1.4788693} .
Cited by: 130; All Open Access, Green Open Access
\end{botherref}
\endbibitem

\bibitem[\protect\citeauthoryear{Hardy}{1919}]{Hardy1919}
\begin{barticle}
\bauthor{\bsnm{Hardy}, \binits{W.B.}}:
\batitle{Iii. the spreading of fluids on glass}.
\bjtitle{The London, Edinburgh, and Dublin Philosophical Magazine and Journal
  of Science}
\bvolume{38}(\bissue{223}),
\bfpage{49}--\blpage{55}
(\byear{1919})
\doiurl{10.1080/14786440708635928}
{\href{https://arxiv.org/abs/https://doi.org/10.1080/14786440708635928}{{https://doi.org/10.1080/14786440708635928}}}
\end{barticle}
\endbibitem

\bibitem[\protect\citeauthoryear{Hygum and
  Popok}{2015}]{c1c7d6568ddf4fa38dd15078b692e6a9}
\begin{bchapter}
\bauthor{\bsnm{Hygum}, \binits{M.}},
\bauthor{\bsnm{Popok}, \binits{V.}}:
\bctitle{Modeling of humidity-related reliability in enclosures with
  electronics}.
In: \beditor{\bsnm{Kutilainen}, \binits{J.}} (ed.)
\bbtitle{IMAPS Nordic Annual Conference 2015 June 8-9, Helsing{\o}r},
pp. \bfpage{106}--\blpage{110}.
\bpublisher{Curran Associates, Inc}, \blocation{???}
(\byear{2015}).
\bcomment{IMAPS Nordic 2015 Conference : International Microelectronics And
  Packaging Society ; Conference date: 01-07-2015}
\end{bchapter}
\endbibitem

\bibitem[\protect\citeauthoryear{ImageJ}{2023}]{imageJ}
\begin{botherref}
\oauthor{\bsnm{ImageJ}}:
{Image Processing and Analysis in Java}
(2023).
\url{https://imagej.net/ij/}
\end{botherref}
\endbibitem

\bibitem[\protect\citeauthoryear{Jacqmin}{2000}]{JACQMIN_2000}
\begin{barticle}
\bauthor{\bsnm{Jacqmin}, \binits{D.}}:
\batitle{Contact-line dynamics of a diffuse fluid interface}.
\bjtitle{Journal of Fluid Mechanics}
\bvolume{402},
\bfpage{57}--\blpage{88}
(\byear{2000})
\doiurl{10.1017/S0022112099006874}
\end{barticle}
\endbibitem

\bibitem[\protect\citeauthoryear{Jiang et~al.}{1979}]{JIANG197974}
\begin{barticle}
\bauthor{\bsnm{Jiang}, \binits{T.-S.}},
\bauthor{\bsnm{Soo-Gun}, \binits{O.H.}},
\bauthor{\bsnm{Slattery}, \binits{J.C.}}:
\batitle{Correlation for dynamic contact angle}.
\bjtitle{Journal of Colloid and Interface Science}
\bvolume{69}(\bissue{1}),
\bfpage{74}--\blpage{77}
(\byear{1979})
\doiurl{10.1016/0021-9797(79)90081-X}
\end{barticle}
\endbibitem

\bibitem[\protect\citeauthoryear{Kubochkin and
  Gambaryan-Roisman}{2022}]{KUBOCHKIN2022101575}
\begin{barticle}
\bauthor{\bsnm{Kubochkin}, \binits{N.}},
\bauthor{\bsnm{Gambaryan-Roisman}, \binits{T.}}:
\batitle{Capillary-driven flow in corner geometries}.
\bjtitle{Current Opinion in Colloid \& Interface Science}
\bvolume{59},
\bfpage{101575}
(\byear{2022})
\doiurl{10.1016/j.cocis.2022.101575}
\end{barticle}
\endbibitem

\bibitem[\protect\citeauthoryear{Lu et~al.}{2016}]{LU201643}
\begin{barticle}
\bauthor{\bsnm{Lu}, \binits{G.}},
\bauthor{\bsnm{Wang}, \binits{X.-D.}},
\bauthor{\bsnm{Duan}, \binits{Y.-Y.}}:
\batitle{A critical review of dynamic wetting by complex fluids: From newtonian
  fluids to non-newtonian fluids and nanofluids}.
\bjtitle{Advances in Colloid and Interface Science}
\bvolume{236},
\bfpage{43}--\blpage{62}
(\byear{2016})
\doiurl{10.1016/j.cis.2016.07.004}
\end{barticle}
\endbibitem

\bibitem[\protect\citeauthoryear{Mohammad~Karim}{2022}]{10.1063/5.0102028}
\begin{barticle}
\bauthor{\bsnm{Mohammad~Karim}, \binits{A.}}:
\batitle{{A review of physics of moving contact line dynamics models and its
  applications in interfacial science}}.
\bjtitle{Journal of Applied Physics}
\bvolume{132}(\bissue{8}),
\bfpage{080701}
(\byear{2022})
\doiurl{10.1063/5.0102028}
{\href{https://arxiv.org/abs/https://pubs.aip.org/aip/jap/article-pdf/doi/10.1063/5.0102028/16511412/080701\_1\_online.pdf}{{https://pubs.aip.org/aip/jap/article-pdf/doi/10.1063/5.0102028/16511412/080701\_1\_online.pdf}}}
\end{barticle}
\endbibitem

\bibitem[\protect\citeauthoryear{Mohammad~Karim
  et~al.}{2016}]{doi:10.1021/acs.langmuir.6b00747}
\begin{barticle}
\bauthor{\bsnm{Mohammad~Karim}, \binits{A.}},
\bauthor{\bsnm{Davis}, \binits{S.H.}},
\bauthor{\bsnm{Kavehpour}, \binits{H.P.}}:
\batitle{Forced versus spontaneous spreading of liquids}.
\bjtitle{Langmuir}
\bvolume{32}(\bissue{40}),
\bfpage{10153}--\blpage{10158}
(\byear{2016})
\doiurl{10.1021/acs.langmuir.6b00747}
{\href{https://arxiv.org/abs/https://doi.org/10.1021/acs.langmuir.6b00747}{{https://doi.org/10.1021/acs.langmuir.6b00747}}}.
\bcomment{PMID: 27643428}
\end{barticle}
\endbibitem

\bibitem[\protect\citeauthoryear{Malkiel and
  Rabinovitch}{2023}]{MALKIEL2023112159}
\begin{barticle}
\bauthor{\bsnm{Malkiel}, \binits{N.}},
\bauthor{\bsnm{Rabinovitch}, \binits{O.}}:
\batitle{Stochastic delamination processes – a comparative analysis}.
\bjtitle{International Journal of Solids and Structures}
\bvolume{267},
\bfpage{112159}
(\byear{2023})
\doiurl{10.1016/j.ijsolstr.2023.112159}
\end{barticle}
\endbibitem

\bibitem[\protect\citeauthoryear{{Mohammad Karim}
  et~al.}{2018}]{MOHAMMADKARIM2018658}
\begin{barticle}
\bauthor{\bsnm{{Mohammad Karim}}, \binits{A.}},
\bauthor{\bsnm{Rothstein}, \binits{J.P.}},
\bauthor{\bsnm{Kavehpour}, \binits{H.P.}}:
\batitle{Experimental study of dynamic contact angles on rough hydrophobic
  surfaces}.
\bjtitle{Journal of Colloid and Interface Science}
\bvolume{513},
\bfpage{658}--\blpage{665}
(\byear{2018})
\doiurl{10.1016/j.jcis.2017.11.075}
\end{barticle}
\endbibitem

\bibitem[\protect\citeauthoryear{Niavarani and
  Priezjev}{2009}]{10.1063/1.3121305}
\begin{barticle}
\bauthor{\bsnm{Niavarani}, \binits{A.}},
\bauthor{\bsnm{Priezjev}, \binits{N.V.}}:
\batitle{{The effective slip length and vortex formation in laminar flow over a
  rough surface}}.
\bjtitle{Physics of Fluids}
\bvolume{21}(\bissue{5}),
\bfpage{052105}
(\byear{2009})
\doiurl{10.1063/1.3121305}
{\href{https://arxiv.org/abs/https://pubs.aip.org/aip/pof/article-pdf/doi/10.1063/1.3121305/14904408/052105\_1\_online.pdf}{{https://pubs.aip.org/aip/pof/article-pdf/doi/10.1063/1.3121305/14904408/052105\_1\_online.pdf}}}
\end{barticle}
\endbibitem

\bibitem[\protect\citeauthoryear{Niavarani and
  Priezjev}{2010}]{PhysRevE.81.011606}
\begin{barticle}
\bauthor{\bsnm{Niavarani}, \binits{A.}},
\bauthor{\bsnm{Priezjev}, \binits{N.V.}}:
\batitle{Modeling the combined effect of surface roughness and shear rate on
  slip flow of simple fluids}.
\bjtitle{Phys. Rev. E}
\bvolume{81},
\bfpage{011606}
(\byear{2010})
\doiurl{10.1103/PhysRevE.81.011606}
\end{barticle}
\endbibitem

\bibitem[\protect\citeauthoryear{Petrov and
  Petrov}{1992}]{doi:10.1021/la00043a013}
\begin{barticle}
\bauthor{\bsnm{Petrov}, \binits{P.}},
\bauthor{\bsnm{Petrov}, \binits{I.}}:
\batitle{A combined molecular-hydrodynamic approach to wetting kinetics}.
\bjtitle{Langmuir}
\bvolume{8}(\bissue{7}),
\bfpage{1762}--\blpage{1767}
(\byear{1992})
\doiurl{10.1021/la00043a013}
{\href{https://arxiv.org/abs/https://doi.org/10.1021/la00043a013}{{https://doi.org/10.1021/la00043a013}}}
\end{barticle}
\endbibitem

\bibitem[\protect\citeauthoryear{Qu{\'e}r{\'e}}{1997}]{Qur1997InertialC}
\begin{barticle}
\bauthor{\bsnm{Qu{\'e}r{\'e}}, \binits{D.}}:
\batitle{Inertial capillarity}.
\bjtitle{EPL (Europhysics Letters)}
\bvolume{39},
\bfpage{533}--\blpage{538}
(\byear{1997})
\end{barticle}
\endbibitem

\bibitem[\protect\citeauthoryear{Ramé}{1997}]{RAME1997245}
\begin{barticle}
\bauthor{\bsnm{Ramé}, \binits{E.}}:
\batitle{The interpretation of dynamic contact angles measured by the wilhelmy
  plate method}.
\bjtitle{Journal of Colloid and Interface Science}
\bvolume{185}(\bissue{1}),
\bfpage{245}--\blpage{251}
(\byear{1997})
\doiurl{10.1006/jcis.1996.4589}
\end{barticle}
\endbibitem

\bibitem[\protect\citeauthoryear{{{Robert Bosch
  GmbH}}}{2024}]{automatedImageAnalysis}
\begin{botherref}
\oauthor{\bsnm{{{Robert Bosch GmbH}}}}:
{BoschResearch/sepMutliphaseFoam, git
  tree/publications/automatedImageAnalysisForInterfaceTrackingInCurvedChannels}.
\url{https://github.com/boschresearch/sepMultiphaseFoam/tree/publications/automatedImageAnalysisForInterfaceTrackingInCurvedChannels}.
Created: 2024-03-14
(2024)
\end{botherref}
\endbibitem

\bibitem[\protect\citeauthoryear{Ramiasa et~al.}{2014}]{RAMIASA2014275}
\begin{barticle}
\bauthor{\bsnm{Ramiasa}, \binits{M.}},
\bauthor{\bsnm{Ralston}, \binits{J.}},
\bauthor{\bsnm{Fetzer}, \binits{R.}},
\bauthor{\bsnm{Sedev}, \binits{R.}}:
\batitle{The influence of topography on dynamic wetting}.
\bjtitle{Advances in Colloid and Interface Science}
\bvolume{206},
\bfpage{275}--\blpage{293}
(\byear{2014})
\doiurl{10.1016/j.cis.2013.04.005} .
\bcomment{Manuel G. Velarde}
\end{barticle}
\endbibitem

\bibitem[\protect\citeauthoryear{Sedev}{2015}]{SEDEV2015661}
\begin{barticle}
\bauthor{\bsnm{Sedev}, \binits{R.}}:
\batitle{The molecular-kinetic approach to wetting dynamics: Achievements and
  limitations}.
\bjtitle{Advances in Colloid and Interface Science}
\bvolume{222},
\bfpage{661}--\blpage{669}
(\byear{2015})
\doiurl{10.1016/j.cis.2014.09.008} .
\bcomment{Reinhard Miller, Honorary Issue}
\end{barticle}
\endbibitem

\bibitem[\protect\citeauthoryear{Shikhmurzaev}{1993}]{SHIKHMURZAEV1993589}
\begin{barticle}
\bauthor{\bsnm{Shikhmurzaev}, \binits{Y.D.}}:
\batitle{The moving contact line on a smooth solid surface}.
\bjtitle{International Journal of Multiphase Flow}
\bvolume{19}(\bissue{4}),
\bfpage{589}--\blpage{610}
(\byear{1993})
\doiurl{10.1016/0301-9322(93)90090-H}
\end{barticle}
\endbibitem

\bibitem[\protect\citeauthoryear{Shikhmurzaev}{2007}]{Shikhmurzaev2007CapillaryFW}
\begin{bchapter}
\bauthor{\bsnm{Shikhmurzaev}, \binits{Y.D.}}:
\bctitle{Capillary flows with forming interfaces}.
(\byear{2007}).
\burl{https://api.semanticscholar.org/CorpusID:118788652}
\end{bchapter}
\endbibitem

\bibitem[\protect\citeauthoryear{Sauer and Kampert}{1998}]{SAUER199828}
\begin{barticle}
\bauthor{\bsnm{Sauer}, \binits{B.B.}},
\bauthor{\bsnm{Kampert}, \binits{W.G.}}:
\batitle{Influence of viscosity on forced and spontaneous spreading: Wilhelmy
  fiber studies including practical methods for rapid viscosity measurement}.
\bjtitle{Journal of Colloid and Interface Science}
\bvolume{199}(\bissue{1}),
\bfpage{28}--\blpage{37}
(\byear{1998})
\doiurl{10.1006/jcis.1997.5319}
\end{barticle}
\endbibitem

\bibitem[\protect\citeauthoryear{Thammanna~Gurumurthy
  et~al.}{2022}]{PhysRevFluids.7.114002}
\begin{barticle}
\bauthor{\bsnm{Thammanna~Gurumurthy}, \binits{V.}},
\bauthor{\bsnm{Baumhauer}, \binits{M.}},
\bauthor{\bsnm{Khair}, \binits{A.}},
\bauthor{\bsnm{Roisman}, \binits{I.V.}},
\bauthor{\bsnm{Tropea}, \binits{C.}},
\bauthor{\bsnm{Garoff}, \binits{S.}}:
\batitle{Forced wetting in a square capillary}.
\bjtitle{Phys. Rev. Fluids}
\bvolume{7},
\bfpage{114002}
(\byear{2022})
\doiurl{10.1103/PhysRevFluids.7.114002}
\end{barticle}
\endbibitem

\bibitem[\protect\citeauthoryear{van~der Waals}{1979}]{van_de_waals}
\begin{barticle}
\bauthor{\bsnm{Waals}, \binits{J.D.}}:
\batitle{The thermodynamic theory of capillarity under the hypothesis of a
  continuous variation of density}.
\bjtitle{Journal of Statistical Physics}
\bvolume{20},
\bfpage{200}--\blpage{244}
(\byear{1979})
\doiurl{10.1007/BF01011514}
\end{barticle}
\endbibitem

\bibitem[\protect\citeauthoryear{Virtanen et~al.}{2020}]{2020SciPy-NMeth}
\begin{barticle}
\bauthor{\bsnm{Virtanen}, \binits{P.}},
\bauthor{\bsnm{Gommers}, \binits{R.}},
\bauthor{\bsnm{Oliphant}, \binits{T.E.}},
\bauthor{\bsnm{Haberland}, \binits{M.}},
\bauthor{\bsnm{Reddy}, \binits{T.}},
\bauthor{\bsnm{Cournapeau}, \binits{D.}},
\bauthor{\bsnm{Burovski}, \binits{E.}},
\bauthor{\bsnm{Peterson}, \binits{P.}},
\bauthor{\bsnm{Weckesser}, \binits{W.}},
\bauthor{\bsnm{Bright}, \binits{J.}},
\bauthor{\bsnm{{van der Walt}}, \binits{S.J.}},
\bauthor{\bsnm{Brett}, \binits{M.}},
\bauthor{\bsnm{Wilson}, \binits{J.}},
\bauthor{\bsnm{Millman}, \binits{K.J.}},
\bauthor{\bsnm{Mayorov}, \binits{N.}},
\bauthor{\bsnm{Nelson}, \binits{A.R.J.}},
\bauthor{\bsnm{Jones}, \binits{E.}},
\bauthor{\bsnm{Kern}, \binits{R.}},
\bauthor{\bsnm{Larson}, \binits{E.}},
\bauthor{\bsnm{Carey}, \binits{C.J.}},
\bauthor{\bsnm{Polat}, \binits{{\. I}.}},
\bauthor{\bsnm{Feng}, \binits{Y.}},
\bauthor{\bsnm{Moore}, \binits{E.W.}},
\bauthor{\bsnm{{VanderPlas}}, \binits{J.}},
\bauthor{\bsnm{Laxalde}, \binits{D.}},
\bauthor{\bsnm{Perktold}, \binits{J.}},
\bauthor{\bsnm{Cimrman}, \binits{R.}},
\bauthor{\bsnm{Henriksen}, \binits{I.}},
\bauthor{\bsnm{Quintero}, \binits{E.A.}},
\bauthor{\bsnm{Harris}, \binits{C.R.}},
\bauthor{\bsnm{Archibald}, \binits{A.M.}},
\bauthor{\bsnm{Ribeiro}, \binits{A.H.}},
\bauthor{\bsnm{Pedregosa}, \binits{F.}},
\bauthor{\bsnm{{van Mulbregt}}, \binits{P.}},
\bauthor{\bsnm{{SciPy 1.0 Contributors}}}:
\batitle{{{SciPy} 1.0: Fundamental Algorithms for Scientific Computing in
  Python}}.
\bjtitle{Nature Methods}
\bvolume{17},
\bfpage{261}--\blpage{272}
(\byear{2020})
\doiurl{10.1038/s41592-019-0686-2}
\end{barticle}
\endbibitem

\bibitem[\protect\citeauthoryear{Voinov}{1976}]{Voinov1976HydrodynamicsOW}
\begin{barticle}
\bauthor{\bsnm{Voinov}, \binits{O.V.}}:
\batitle{Hydrodynamics of wetting}.
\bjtitle{Fluid Dynamics}
\bvolume{11},
\bfpage{714}--\blpage{721}
(\byear{1976})
\end{barticle}
\endbibitem

\bibitem[\protect\citeauthoryear{Vega et~al.}{2005}]{doi:10.1021/la051341z}
\begin{barticle}
\bauthor{\bsnm{Vega}, \binits{M.-J.}},
\bauthor{\bsnm{Seveno}, \binits{D.}},
\bauthor{\bsnm{Lemaur}, \binits{G.}},
\bauthor{\bsnm{Adão}, \binits{M.-H.}},
\bauthor{\bsnm{De~Coninck}, \binits{J.}}:
\batitle{Dynamics of the rise around a fiber: Experimental evidence of the
  existence of several time scales}.
\bjtitle{Langmuir}
\bvolume{21}(\bissue{21}),
\bfpage{9584}--\blpage{9590}
(\byear{2005})
\doiurl{10.1021/la051341z}
{\href{https://arxiv.org/abs/https://doi.org/10.1021/la051341z}{{https://doi.org/10.1021/la051341z}}}.
\bcomment{PMID: 16207039}
\end{barticle}
\endbibitem

\bibitem[\protect\citeauthoryear{Wang et~al.}{2013}]{WANG6532474}
\begin{barticle}
\bauthor{\bsnm{Wang}, \binits{H.}},
\bauthor{\bsnm{Liserre}, \binits{M.}},
\bauthor{\bsnm{Blaabjerg}, \binits{F.}}:
\batitle{Toward reliable power electronics: Challenges, design tools, and
  opportunities}.
\bjtitle{IEEE Industrial Electronics Magazine}
\bvolume{7}(\bissue{2}),
\bfpage{17}--\blpage{26}
(\byear{2013})
\doiurl{10.1109/MIE.2013.2252958}
\end{barticle}
\endbibitem

\bibitem[\protect\citeauthoryear{Wong et~al.}{1992}]{WONG1992317}
\begin{barticle}
\bauthor{\bsnm{Wong}, \binits{H.}},
\bauthor{\bsnm{Morris}, \binits{S.}},
\bauthor{\bsnm{Radke}, \binits{C.J.}}:
\batitle{Three-dimensional menisci in polygonal capillaries}.
\bjtitle{Journal of Colloid and Interface Science}
\bvolume{148}(\bissue{2}),
\bfpage{317}--\blpage{336}
(\byear{1992})
\doiurl{10.1016/0021-9797(92)90171-H}
\end{barticle}
\endbibitem

\bibitem[\protect\citeauthoryear{Wang
  et~al.}{2016}]{doi:10.1021/acs.langmuir.6b01357}
\begin{barticle}
\bauthor{\bsnm{Wang}, \binits{X.}},
\bauthor{\bsnm{Venzmer}, \binits{J.}},
\bauthor{\bsnm{Bonaccurso}, \binits{E.}}:
\batitle{Surfactant-enhanced spreading of sessile water drops on polypropylene
  surfaces}.
\bjtitle{Langmuir}
\bvolume{32}(\bissue{33}),
\bfpage{8322}--\blpage{8328}
(\byear{2016})
\doiurl{10.1021/acs.langmuir.6b01357}
{\href{https://arxiv.org/abs/https://doi.org/10.1021/acs.langmuir.6b01357}{{https://doi.org/10.1021/acs.langmuir.6b01357}}}.
\bcomment{PMID: 27448154}
\end{barticle}
\endbibitem

\bibitem[\protect\citeauthoryear{Yang et~al.}{2020}]{YANG202021}
\begin{barticle}
\bauthor{\bsnm{Yang}, \binits{L.}},
\bauthor{\bsnm{Chen}, \binits{C.}},
\bauthor{\bsnm{Hu}, \binits{Y.}},
\bauthor{\bsnm{Wei}, \binits{F.}},
\bauthor{\bsnm{Cui}, \binits{J.}},
\bauthor{\bsnm{Zhao}, \binits{Y.}},
\bauthor{\bsnm{Xu}, \binits{X.}},
\bauthor{\bsnm{Chen}, \binits{X.}},
\bauthor{\bsnm{Sun}, \binits{D.}}:
\batitle{Three-dimensional bacterial cellulose/polydopamine/tio2 nanocomposite
  membrane with enhanced adsorption and photocatalytic degradation for dyes
  under ultraviolet-visible irradiation}.
\bjtitle{Journal of Colloid and Interface Science}
\bvolume{562},
\bfpage{21}--\blpage{28}
(\byear{2020})
\doiurl{10.1016/j.jcis.2019.12.013}
\end{barticle}
\endbibitem

\bibitem[\protect\citeauthoryear{Yuan and Zhao}{2010}]{PhysRevLett.104.246101}
\begin{barticle}
\bauthor{\bsnm{Yuan}, \binits{Q.}},
\bauthor{\bsnm{Zhao}, \binits{Y.-P.}}:
\batitle{Precursor film in dynamic wetting, electrowetting, and
  electro-elasto-capillarity}.
\bjtitle{Phys. Rev. Lett.}
\bvolume{104},
\bfpage{246101}
(\byear{2010})
\doiurl{10.1103/PhysRevLett.104.246101}
\end{barticle}
\endbibitem

\bibitem[\protect\citeauthoryear{Yue et~al.}{2010}]{YUE_ZHOU_FENG_2010}
\begin{barticle}
\bauthor{\bsnm{Yue}, \binits{P.}},
\bauthor{\bsnm{Zhou}, \binits{C.}},
\bauthor{\bsnm{Feng}, \binits{J.J.}}:
\batitle{Sharp-interface limit of the cahn–hilliard model for moving contact
  lines}.
\bjtitle{Journal of Fluid Mechanics}
\bvolume{645},
\bfpage{279}--\blpage{294}
(\byear{2010})
\doiurl{10.1017/S0022112009992679}
\end{barticle}
\endbibitem

\bibitem[\protect\citeauthoryear{Zhang et~al.}{2023}]{ZHANG2023102861}
\begin{barticle}
\bauthor{\bsnm{Zhang}, \binits{Y.}},
\bauthor{\bsnm{Guo}, \binits{M.}},
\bauthor{\bsnm{Seveno}, \binits{D.}},
\bauthor{\bsnm{{De Coninck}}, \binits{J.}}:
\batitle{Dynamic wetting of various liquids: Theoretical models, experiments,
  simulations and applications}.
\bjtitle{Advances in Colloid and Interface Science}
\bvolume{313},
\bfpage{102861}
(\byear{2023})
\doiurl{10.1016/j.cis.2023.102861}
\end{barticle}
\endbibitem

\bibitem[\protect\citeauthoryear{Zhang et~al.}{2024}]{zhang_published_2024}
\begin{botherref}
\oauthor{\bsnm{Zhang}, \binits{H.}},
\oauthor{\bsnm{Lippert}, \binits{A.}},
\oauthor{\bsnm{Leonhardt}, \binits{R.}},
\oauthor{\bsnm{Tolle}, \binits{T.}},
\oauthor{\bsnm{Nagel}, \binits{L.}},
\oauthor{\bsnm{Fricke}, \binits{M.}},
\oauthor{\bsnm{Maric}, \binits{T.}}:
Experimental study of dynamic wetting behavior through curved microchannels
  with automated image analysis.
Experiments in Fluids
\textbf{65}(95)
(2024)
\doiurl{10.1007/s00348-024-03828-7}
\end{botherref}
\endbibitem

\bibitem[\protect\citeauthoryear{Zhang et~al.}{2024-03-13}]{TUDatalib}
\begin{botherref}
\oauthor{\bsnm{Zhang}, \binits{H.}},
\oauthor{\bsnm{Lippert}, \binits{A.}},
\oauthor{\bsnm{Leonhardt}, \binits{R.}},
\oauthor{\bsnm{Tolle}, \binits{T.}},
\oauthor{\bsnm{Nagel}, \binits{L.}},
\oauthor{\bsnm{Fricke}, \binits{M.}},
\oauthor{\bsnm{Maric}, \binits{T.}}:
Experimental study of dynamic wetting behavior through curved microchannels
  with automated image analysis - Data.
Technical University of Darmstadt
(2024-03-13).
\doiurl{10.48328/tudatalib-1382} .
\url{https://tudatalib.ulb.tu-darmstadt.de/handle/tudatalib/4176}
\end{botherref}
\endbibitem

\end{thebibliography}

\end{document}